\documentclass[12pt]{article}
\usepackage{amsmath}
\usepackage{graphicx}
\usepackage{amsfonts}
\usepackage{amssymb}
\newtheorem{theorem}{Theorem}

\newtheorem{definition}[theorem]{Definition}
\newtheorem{example}[theorem]{Example}

\newtheorem{remark}[theorem]{Remark}

\begin{document}

\title{A Quick Glance at Quantum Cryptography}
\author{Samuel J. Lomonaco, Jr.\thanks{Partially supported by ARL Contract
\#DAAL01-95-P-1884, ARO Grant \#P-38804-PH-QC, and the L-O-O-P Fund.}\\Dept. of Comp. Sci. \& Elect. Engr.\\University of Maryland Baltimore County\\1000 Hilltop Circle\\Baltimore, MD 21250\\E-Mail: Lomonaco@UMBC.EDU\\WebPage: http://www.csee.umbc.edu/\symbol{126}lomonaco}
\date{November 8, 1998}
\maketitle
\begin{abstract}
The recent application of the principles of quantum mechanics to cryptography
has led to a remarkable new dimension in secret communication. As a result of
these new developments, it is now possible to construct cryptographic
communication systems which detect unauthorized eavesdropping should it occur,
and which give a guarantee of no eavesdropping should it not occur.
\end{abstract}
\tableofcontents

\section{Cryptographic systems before quantum cryptography}

A brief description of a classical cryptographic system (CCS) \cite{Shannon1}
is illustrated in Fig. 1.
\begin{center}
\includegraphics[
height=4.5602in,
width=6.2068in
]%
{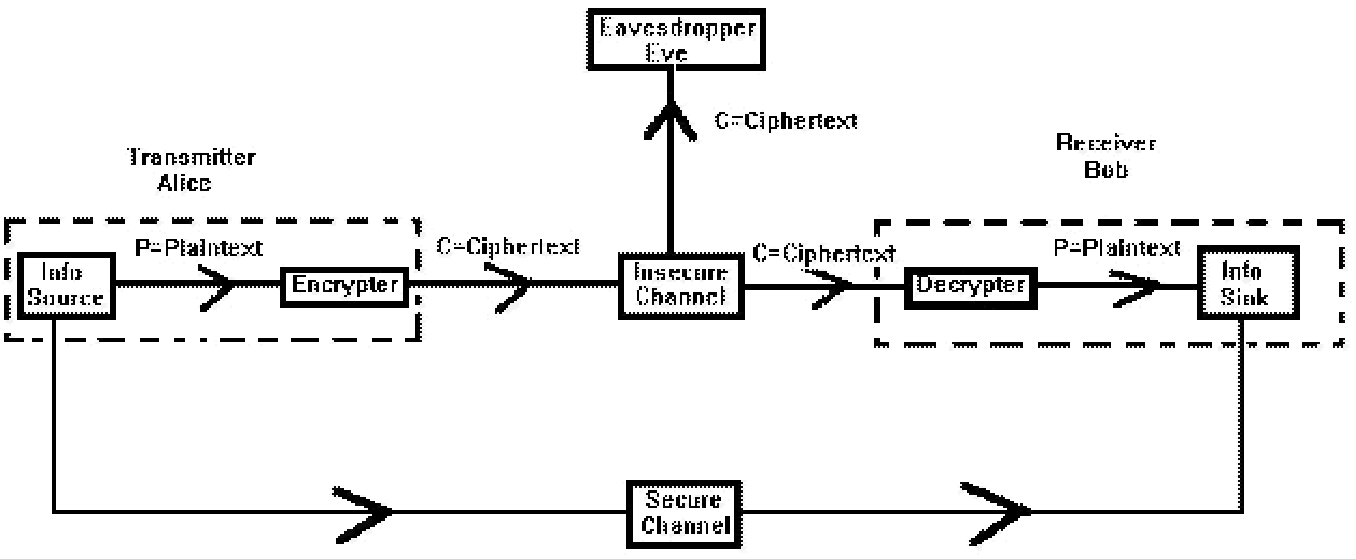}%
\\
Figure 1. A classical cryptographic communication system.
\end{center}

A message, called \textbf{plaintext} $P$, is encrypted via a secret
\textbf{key} $K$ into \textbf{ciphertext} $C$, sent over a non-secure
communication channel, and finally decrypted via a secret \textbf{key}
$K^{\prime}$ back into readable plaintext $P$. Following the conventions of
the cryptographic literature, we will refer to the transmitter as
\textbf{Alice}, to the receiver as \textbf{Bob}, and to an adversarial
eavesdropper as \textbf{Eve}.

There are classical cryptographic systems which are \textbf{perfectly secure}
(see \cite{Shannon1}), such as the \textbf{Vernam cipher}, better know as the
\textbf{one time pad}, which uses a perfectly random key $K$ equal in length
to the length of the message. The chief practical difficulty with such
perfectly secure systems is that Alice must first communicate a random key in
secret via some totally secure channel. In most cases, the length of the key
makes this secure communication impractical and too costly. Because of the
large cost of transmitting such long keys over a secure channel, Alice is
frequently tempted to use the same key twice. If she makes this fatal mistake,
then her ciphertext immediately changes from being perfectly secure to
ciphertext that is easily read by Eve.

Thus, for almost all practical cryptographic systems, the key $K$ is
substantially shorter than the length of the plaintext. As a result, the
ciphertext is no longer perfectly secure. However, if the encryption method
and key $K$ are wisely chosen, then Alice's communication to Bob will be
\textbf{practically secure}. By ``practically secure,'' we mean that, although
adversary Eve is theoretically able to decrypt Alice and Bob's communication
without any knowledge of their key, she can not do so because the required
computational time and resources are simply beyond her capability and means.
The \textbf{Data Encryption Standard (DES)} is believed to be an example of
such a practically secure encryption system. (See for example \cite{Stinson1}.)

In any case, one Achilles heal of classical cryptographic communication
systems is that secret communication can only take place after a key is
communicated in secret over a totally secure communication channel. This is
frequently referred to as the ``\textsc{catch 22}'' of cryptography, i.e.,

\bigskip\ 

\noindent\textbf{Catch 22}: Before Alice and Bob can communicate in secret,
they must first communicate in secret.

\bigskip\ 

There is even more to this catch 22, namely:

\bigskip\ 

\noindent\textbf{Catch 22a}: Even if Alice and Bob somehow succeed in
communicating their key over a secure communication channel, there is simply
no classical cryptographic mechanism guaranteeing with total certainty that
their key was transmitted securely, i.e., that their ``secure'' communication
channel is free of Eve's unauthorized intrusion.

\bigskip\ 

As we shall see, quantum encryption does provide a means of circumventing this
impasse of intrusion detection.

\bigskip\ 

A proposed solution to the catch 22 of classical cryptographic communication
systems is the modern \textbf{public key cryptographic system} (\textbf{PKCS})
as illustrated in Fig. 2. (See \cite{Diffie1} \cite{Diffie2}.)

For public key cryptographic systems, it is no longer necessary for Alice and
Bob to exchange key over a secure channel. Instead, Alice and Bob both create
their own individual encryption/decryption key pairs $(E_{A},D_{A})$ and
$(E_{B},D_{B})$, respectively. Then they both keep their decryption keys
$D_{A}$ and $D_{B}$ secret from everyone, including each other, and
``publish'' or publicly broadcast their encryption keys $E_{A}$ and $E_{B}$
for the entire world to see. The security of such a public key cryptographic
system depends on the selection of an encryption/decryption algorithm which is
a \textbf{trapdoor function}. As a result, recovering the decryption key from
the encryption key is computationally infeasible. The RSA public key
cryptographic system is believed to be an example of such a cryptographic
system. (See for example \cite{Stinson1}.)%

\begin{center}
\includegraphics[
height=3.0312in,
width=4.3215in
]%
{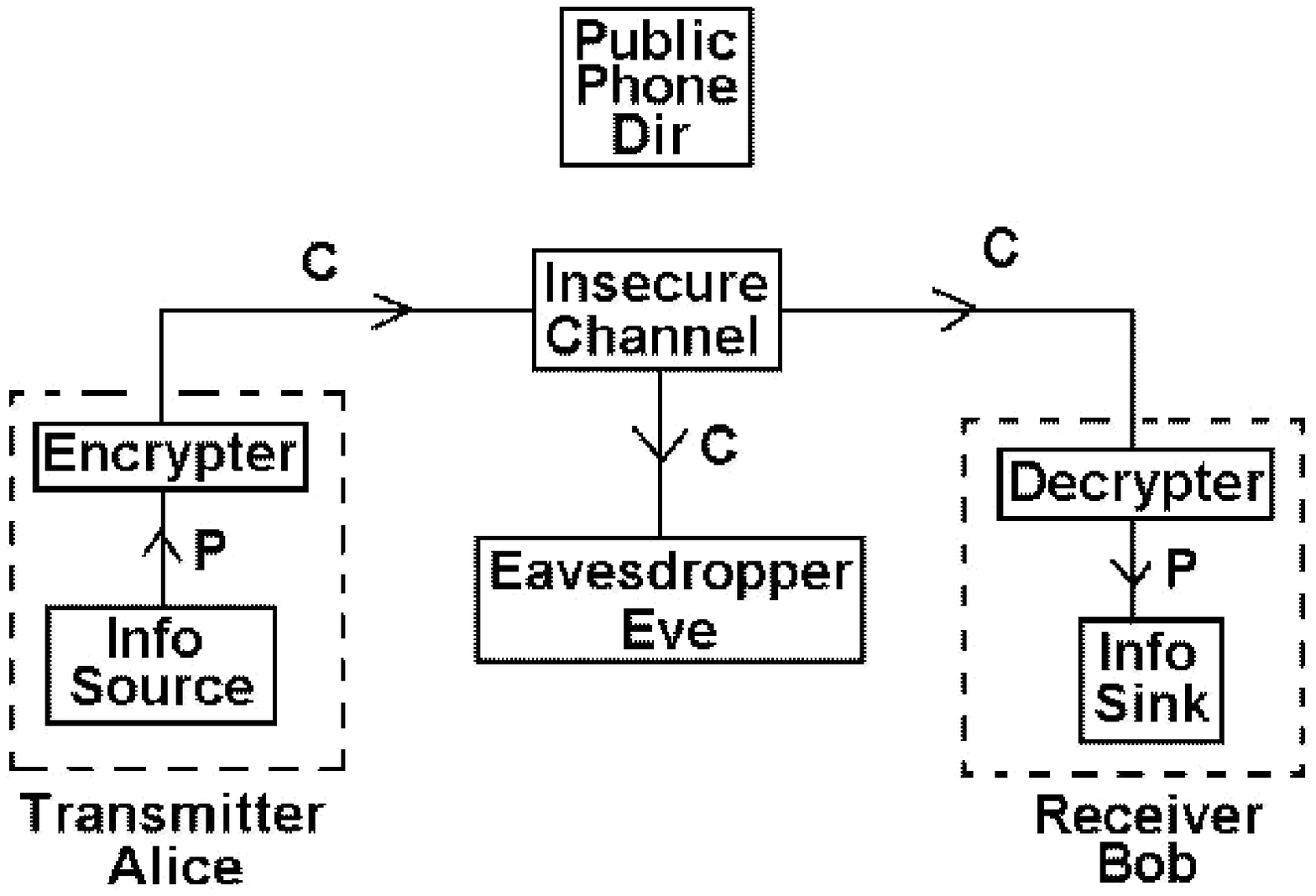}%
\\
Figure 2. A public key cryptographic communication system.
\end{center}

One major drawback to public key cryptographic systems is that no one has yet
been able to prove that practical trapdoor functions exist. As a result, no
one is really sure how secure such public key cryptographic systems are.
Moreover, if researchers succeed in building a feasible quantum computer,
Shor's quantum factoring algorithm \cite{Shor1a} could break RSA easily, i.e.,
in polynomial time.

Yet another drawback to public key cryptographic systems is that, in terms of
some everyday implementations, such systems frequently do not circumvent the
catch 22 of classical cryptography after all. The keys for many practical
public key cryptographic systems are frequently managed by a \textbf{key bank}
that is independent of Alice and Bob. Thus, secret communications over a
secure channel from the key bank to Alice and Bob are required before Alice
and Bob can secretly communicate.

\bigskip

Finally, it should be noted that the most important contribution of quantum
cryptography is a mechanism for detecting eavesdropping. This is a totally new
contribution to the field of cryptography. Neither classical cryptographic
systems nor public key cryptographic systems have such a capability. In the
next section, we will see how quantum mechanics provides a means for detecting intrusion.

\bigskip\ 

\section{Preamble to quantum cryptography}

The recent results in quantum cryptography are based on the \textbf{Heisenberg
uncertainty principle} of quantum mechanics\footnote{For those not familiar
with quantum mechanics, please refer to appendix C for a quick overview.}.
Using standard Dirac notation\footnote{As outlined in Appendix C}, this
principle can be succinctly stated as follows:

\begin{description}
\item \textbf{Heisenberg Uncertainty Principle:} For any two quantum
mechanical \textbf{observables} $A$ and $B$
\[
\left\langle \left(  \Delta A\right)  ^{2}\right\rangle \left\langle \left(
\Delta B\right)  ^{2}\right\rangle \geq\frac{1}{4}\left\|  \left\langle
\left[  A,B\right]  \right\rangle \right\|  ^{2}\text{,}%
\]
where
\[
\Delta A=A-\left\langle A\right\rangle \text{\qquad and\qquad}\Delta
B=B-\left\langle B\right\rangle \text{,}%
\]
and where
\[
\left[  A,B\right]  =AB-BA\text{.}%
\]
\end{description}

Thus, $\left\langle \left(  \Delta A\right)  ^{2}\right\rangle $ and
$\left\langle \left(  \Delta B\right)  ^{2}\right\rangle $ are variances which
measure the uncertainty of observables $A$ and $B$. For \textbf{incompatible}
observables, i.e., for observables $A$ and $B$ such that $\left[  A,B\right]
\neq0$, reducing the uncertainty $\left\langle \left(  \Delta A\right)
^{2}\right\rangle $ of $A$ forces the uncertainty $\left\langle \left(  \Delta
B\right)  ^{2}\right\rangle $ of $B$ to increase, and vice versa. Thus the
observables $A$ and $B$ can not be simultaneously measured to arbitrary
precision. Measuring one of the observables interferes with the measurement of
the other.

\bigskip\ 

Young's double slit experiment is an example suggesting how Heisenberg's
uncertainty principle could be used for detecting eavesdropping in a
cryptographic communications. This experiment is illustrated in Fig. 3.%

\begin{center}
\includegraphics[
height=3.1964in,
width=5.3471in
]%
{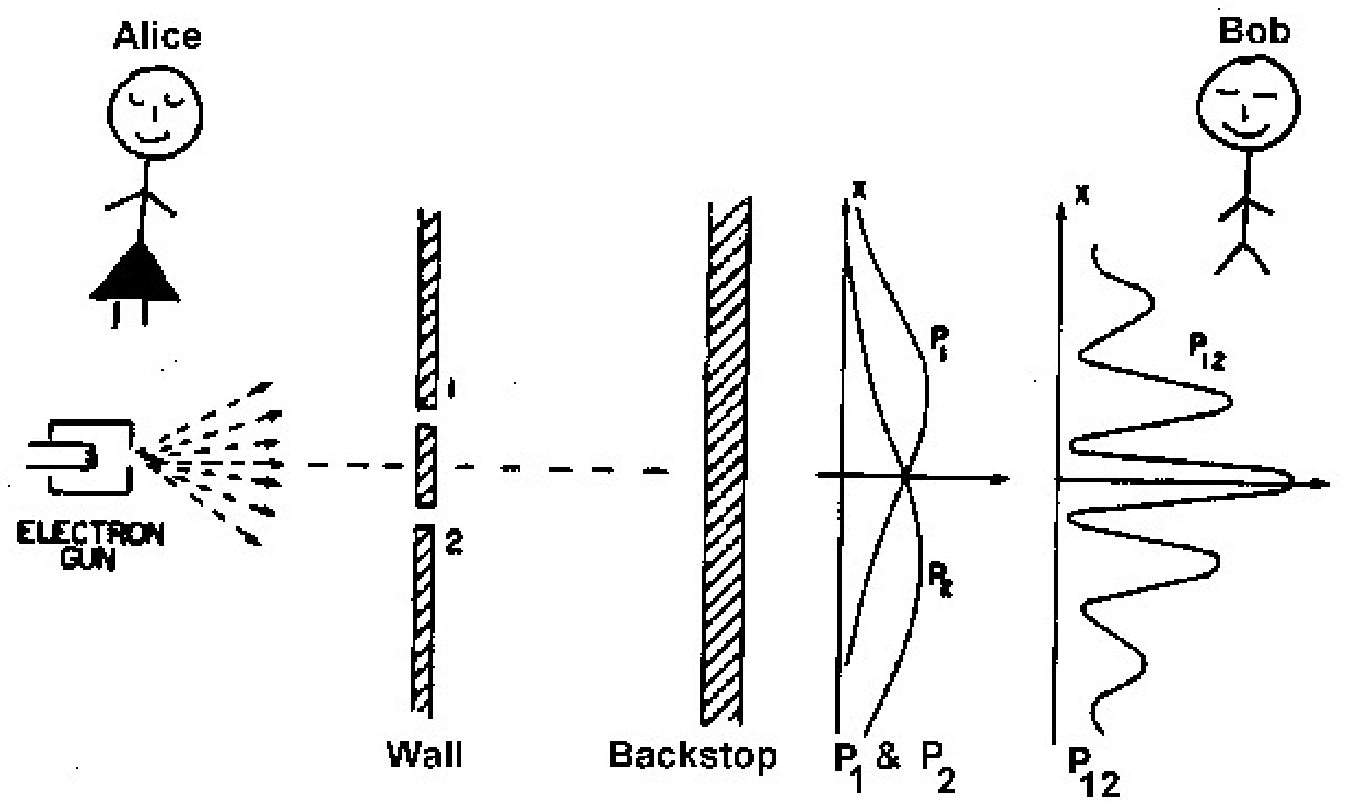}%
\\
Figure 3. Young's double slit experiment when electron trajectories are not
observed. The first of two incompatible observables is measured.
\end{center}

\smallskip

An electron gun randomly emits electrons over a fairly large angular spread.
In front of the gun is a metal wall with two small slits. Beyond the wall is a
backstop that absorbs the electrons that pass through the two slits. The
probability density pattern of the absorbed electrons is described by the
curves $P_{1}$, $P_{2}$, and $P_{21}$ which, for the convenience of the
reader, have been drawn behind the backstop. The curve $P_{1}$ denotes the
probability density pattern if only slit 1 is open. The curve $P_{2}$ denotes
the probability density pattern if only slit 2 is open. Finally, the curve
$P_{12}$ denotes the probability density pattern if both slits 1 and 2 are
open. Thus, $P_{12}$ shows a quantum mechanical interference pattern
demonstrating the wave nature of electrons.%

\begin{center}
\includegraphics[
height=3.3754in,
width=5.3549in
]%
{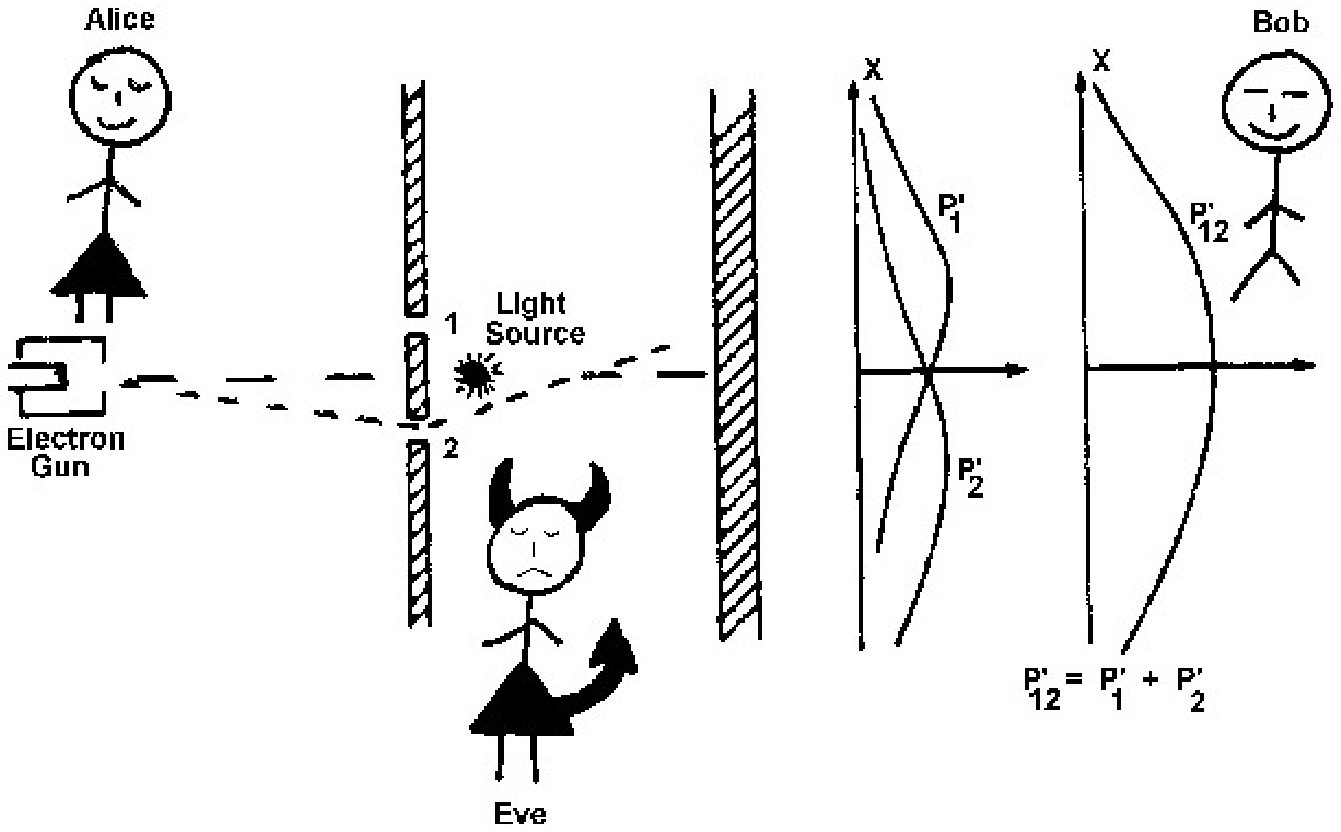}%
\\
Figure 4. Young's double slit experiment when electron trajectories are
observed by Eve. The second of two incompatible observables is measured.
\end{center}

Comparing this with our description of a classical cryptographic system, the
electron gun can be thought of as the transmitter Alice. And the interference
pattern $P_{12}$ can be thought of as the message received by Bob. If however,
Eve tries to eavesdrop by trying to detect through which slit each electron
passes, as illustrated in Fig. 4, the interference pattern $P_{12}$ is
destroyed and replaced by the bell curve $P_{12}^{\prime} $ (which is a
classical superposition of curves $P_{1}^{\prime}$ and $P_{2}^{\prime}$) drawn
in Fig. 4, thus demonstrating the particle nature of the electron. As a
result, Bob knows with certainty that Eve is eavesdropping in on his
communication with Alice. Bob knows that, because of the Heisenberg
uncertainty principle, both the wave and particle natures of the electron can
not be simultaneously detected.

\bigskip\ 

In the next sections, we describe a number of methods, i.e., \textbf{quantum
cryptographic communication protocols}, that utilize the Heisenberg
uncertainty principle to communicate random binary sequences (i.e., keys) with
automatic eavesdrop detection. These quantum communication protocols provide a
means of circumventing the ``catch 22'' of classical cryptographic systems. As
a result, the perfect security of the Vernam cipher (i.e., one-time-pad) is an
inexpensively implementable reality.

All the quantum cryptographic systems we discuss in this paper can be
implemented by transmissions over fiber optic cable of individual photons,
each with a single bit encoded in its quantum mechanical state space. We
describe all of these systems in terms of the polarization states of a single
photon. It should be noted that they could equally well be described in terms
of any two-state quantum system. Examples of such a system include a
spin-$\frac{1}{2}$ particle, and a two-state level atom.

The quantum cryptographic protocols discussed will of necessity use some
encoding scheme (or schemes) which associates the bits $0$ and $1$ with
distinct quantum states. We call such an association a \textbf{quantum
alphabet}. Should the associated states be orthogonal, we call the encoding
scheme an \textbf{orthogonal quantum alphabet}.

\bigskip\ 

\section{The BB84 quantum cryptographic protocol without noise}

The first quantum cryptographic communication protocol, called \textbf{BB84},
was invented in 1984 by Bennett and Brassard \cite{Bennett5}\footnote{Quantum
cryptographic protocols evolved from the earlier work of Wiesner
\cite{Wiesner1}.}. This protocol has been experimentally demonstrated to work
for a transmission over 30 km of fiber optic cable \cite{Phoenix2}
\cite{Townsend1} \cite{Townsend2} \cite{Townsend3}, and also over free space
for a distance of over one hundred meters\cite{Jacobs1} \cite{Franson2}. It is
speculated, but not yet experimentally verified, that the BB84 protocol should
be implementable over distances of at least 100 km.

In this section we describe the BB84 protocol in a noise free environment. In
the next section, we extend the protocol to one in which noise is
considered.\footnote{The proofs given in this and the next section are based
on the assumption that Eve uses the opaque eavedropping strategy. \ Other
eavesdropping strategies are briefly discussed in section 8 of this paper.}

We now describe the BB84 protocol in terms of the polarization states of a
single photon. Please note that the BB84 protocol could equally well be
described in terms of any other two-state quantum system.

\bigskip

Let $\mathcal{H}$ be the two dimensional Hilbert space whose elements
representate the polarization states of a single photon. In describing BB84,
we use two different orthogonal bases of $\mathcal{H}$. They are the
\textbf{circular polarization basis}, which consists of the kets
\[
\left|  \curvearrowright\right\rangle \text{ and }\left|  \curvearrowleft
\right\rangle
\]
for \textbf{right} and \textbf{left circular polarization states},
respectively, and the \textbf{linear polarization basis} which consists of the
kets
\[
\left|  \updownarrow\right\rangle \text{ and }\left|  \leftrightarrow
\right\rangle
\]
for \textbf{vertical} and \textbf{horizontal linear polarization states}, respectively.

The BB84 protocol utilizes any two incompatible orthogonal quantum alphabets
in the Hilbert space $\mathcal{H}$. For our description of BB84, we have
selected the \textbf{circular polarization quantum alphabet} $\mathcal{A}%
_{\odot}$
\[%
\begin{tabular}
[c]{c||c}%
Symbol & Bit\\\hline\hline
$\left|  \curvearrowright\right\rangle $ & $1$\\\hline
$\left|  \curvearrowleft\right\rangle $ & $0$\\\hline
& \\
\multicolumn{2}{c}{Circular Polarization}\\
\multicolumn{2}{c}{Quantum Alphabet $\mathcal{A}_{\odot}$}%
\end{tabular}
\]
and the \textbf{linear polarization quantum alphabet} $\mathcal{A}_{\boxplus
}$
\[%
\begin{tabular}
[c]{c||c}%
Symbol & Bit\\\hline\hline
$\left|  \updownarrow\right\rangle $ & $1$\\\hline
$\left|  \leftrightarrow\right\rangle $ & $0$\\\hline
& \\
\multicolumn{2}{c}{Linear Polarization}\\
\multicolumn{2}{c}{Quantum Alphabet $\mathcal{A}_{\boxplus}$}%
\end{tabular}
\]

\bigskip\ 

Bennett and Brassard note that, if Alice were to use only one specific
orthogonal quantum alphabet for her communication to Bob, then Eve's
eavesdropping could go undetected. For Eve could intercept Alice's
transmission with 100\% accuracy, and then imitate Alice by retransmitting her
measurements to Bob. If, for example, Alice used only the orthogonal quantum
alphabet $\mathcal{A}_{\odot}$, then Eve could measure each bit of Alice's
transmission with a device based on some circular polarization measurement
operator such as
\[
\left|  \curvearrowright\right\rangle \left\langle \curvearrowright\right|
\text{\qquad or\qquad}\left|  \curvearrowleft\right\rangle \left\langle
\curvearrowleft\right|
\]
Or if, Alice used only the orthogonal quantum alphabet $\mathcal{A}_{\boxplus
}$, then Eve could measure each transmitted bit with a device based on some
linear polarization measurement operator such as
\[
\left|  \updownarrow\right\rangle \left\langle \updownarrow\right|
\text{\qquad or\qquad}\left|  \leftrightarrow\right\rangle \left\langle
\leftrightarrow\right|
\]
The above strategy used by Eve is called \textbf{opaque eavesdropping}
\cite{Ekert1}. (We will consider other and more sophisticated eavesdropping
strategies later.)

\bigskip\ 

To assure the detection of Eve's eavesdropping, Bennett and Brassard require
Alice and Bob to communicate in two stages, the first stage over a one-way
quantum communication channel from Alice to Bob, the second stage over a
two-way public communication channel. (Please refer to Figure 5.)

\bigskip\ 

\subsection{Stage 1. Communication over a quantum channel}

\bigskip\ 

In the first stage, Alice is required, each time she transmits a single bit,
to use randomly with equal probability one of the two orthogonal alphabets
$\mathcal{A}_{\odot}$ or $\mathcal{A}_{\boxplus}$. Since no measurement
operator of $\mathcal{A}_{\odot}$ is compatible with any measurement operator
of $\mathcal{A}_{\boxplus}$, it follows from the Heisenberg uncertainty
principle that no one, not even Bob or Eve, can receive Alice's transmission
with an accuracy greater than 75\%.%

\begin{center}
\includegraphics[
height=3.9124in,
width=5.214in
]%
{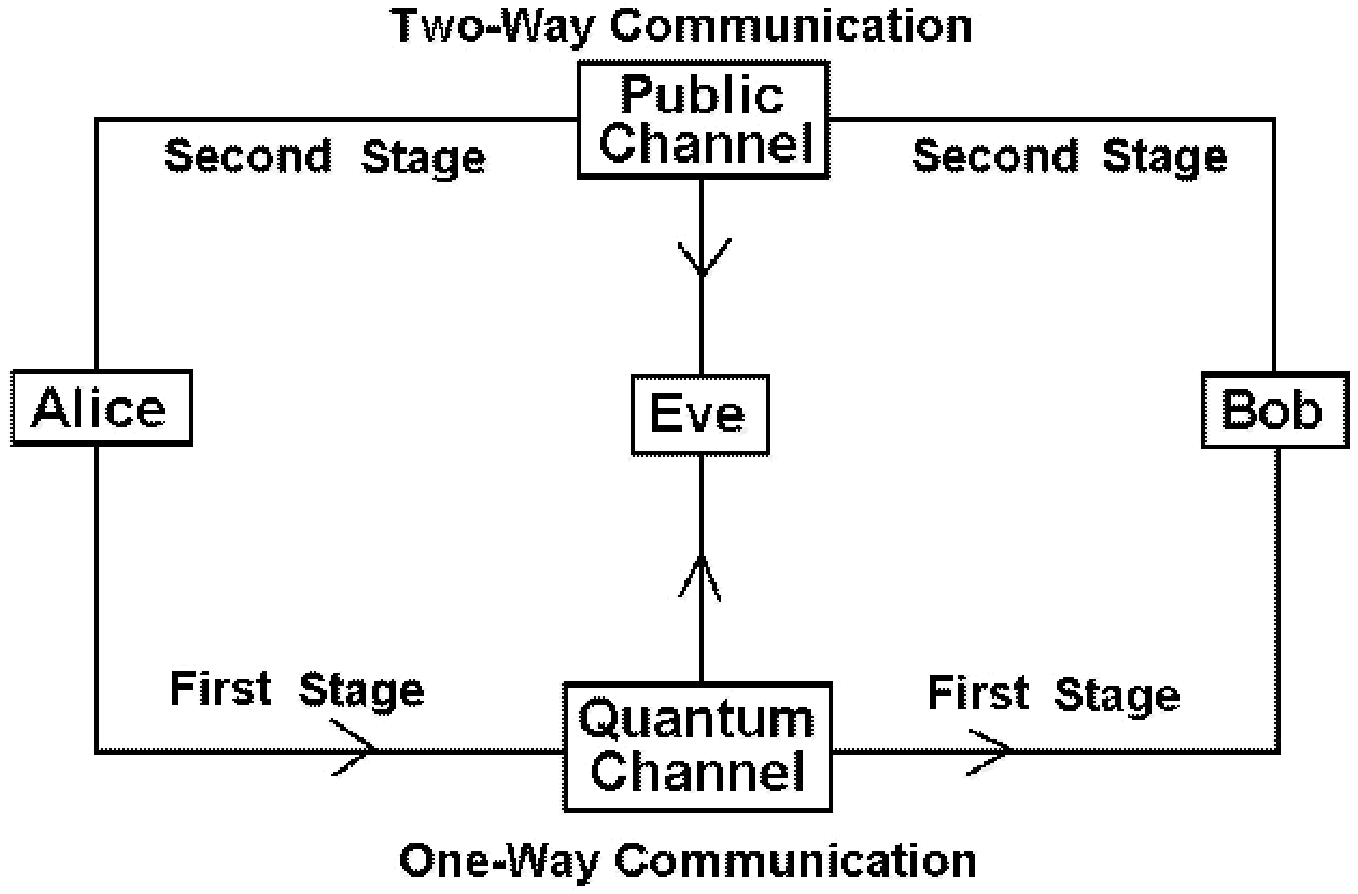}%
\\
Figure 5. A quantum cryptographic communication system for securely
transfering random key.
\end{center}

This can be seen as follows. For each bit transmitted by Alice, one can choose
a measurement operator compatible with either $\mathcal{A}_{\odot}$ or
$\mathcal{A}_{\boxplus}$, but not both. Because of incompatibility, there is
no simultaneous measurement operator for both $\mathcal{A}_{\odot}$ and
$\mathcal{A}_{\boxplus}$. Since one has no knowledge of Alice's secret choice
of quantum alphabet, 50\% of the time (i.e., with probability $\frac12$) one
will guess correctly, i.e., choose a measurement operator compatible with
Alice's choice, and 50\% of the time (i.e., with probability $\frac12$) one
will guess incorrectly. If one guesses correctly, then Alice's transmitted bit
is received with probability $1$. On the other hand, if one guesses
incorrectly, then Alice's transmitted bit is received correctly with
probability $\frac12$. Thus in general, the probability of correctly receiving
Alice's transmitted bit is
\[
P=\frac12\cdot1+\frac12\cdot\frac12=\frac34
\]

\bigskip\ 

For each bit transmitted by Alice, we assume that Eve performs one of two
actions, opaque eavesdropping with probability $\lambda$, $0\leq\lambda\leq1$,
or no eavesdropping with probability $1-\lambda$. Thus, if $\lambda=1$, Eve is
eavesdropping on each transmitted bit; and if $\lambda=0 $, Eve is not
eavesdropping at all.

\bigskip\ 

Because Bob's and Eve's choice of measurement operators are stochastically
independent of each other and of Alice's choice of alphabet, Eve's
eavesdropping has an immediate and detectable impact on Bob's received bits.
Eve's eavesdropping causes Bob's error rate to jump from $\frac14$ to
\[
\frac14(1-\lambda)+\frac38\lambda=\frac14+\frac\lambda8
\]
Thus, if Eve eavesdrops on every bit, i.e., if $\lambda=1$, then Bob's error
rate jumps from $\frac14$ to $\frac38$, a 50\% increase.

\subsection{Stage 2. Communication in two phases over a public channel}

\bigskip\ 

In stage 2, Alice and Bob communicate in two phases over a public channel to
check for Eve's presence by analyzing Bob's error rate.

\subsubsection{Phase 1 of Stage 2. Extraction of raw key}

\bigskip\ 

Phase 1 of stage 2 is dedicated to eliminating the bit locations (and hence
the bits at these locations) at which error could have occurred without Eves
eavesdropping. Bob begins by publicly communicating to Alice which measurement
operators he used for each of the received bits. Alice then in turn publicly
communicates to Bob which of his measurement operator choices were correct.
After this two way communication, Alice and Bob delete the bits corresponding
to the incompatible measurement choices to produce shorter sequences of bits
which we call respectively \textbf{Alice's raw key} and \textbf{Bob's raw key}.

If there is no intrusion, then Alice's and Bob's raw keys will be in total
agreement. However, if Eve has been at work, then corresponding bits of
Alice's and Bob's raw keys will not agree with probability
\[
0\cdot(1-\lambda)+\frac14\cdot\lambda=\frac\lambda4
\]

\subsubsection{Phase 2 of Stage 2. Detection of Eve's intrusion via error detection}

\bigskip\ 

Alice and Bob now initiate a two way conversation over the public channel to
test for Eve's presence.

In the absence of noise, any discrepancy between Alice's and Bob's raw keys is
proof of Eve's intrusion. So to detect Eve, Alice and Bob select a publicly
agreed upon random subset of $m$ bit locations in the raw key, and publicly
compare corresponding bits, making sure to discard from raw key each bit as it
is revealed.

Should at least one comparison reveal an inconsistency, then Eve's
eavesdropping has been detected, in which case Alice and Bob return to stage 1
and start over. On the other hand, if no inconsistencies are uncovered, then
the probability that Eve escapes detection is:
\[
P_{false}=\left(  1-\frac\lambda4\right)  ^{m}%
\]
For example, if $\lambda=1$ and $m=200$, then
\[
P_{false}=\left(  \frac34\right)  ^{200}\approx10^{-25}%
\]
Thus, if $P_{false}$ is sufficiently small, Alice and Bob agree that Eve has
not eavesdropped, and accordingly adopt the remnant raw key as their
\textbf{final secret key}.

\bigskip\ 

\section{The BB84 quantum cryptographic protocol with noise}

In this section, the BB84 protocol is extended to a noisy environment. Since,
in a noisy environment, Alice and Bob can not distinguish between error caused
by noise and error caused by Eve's eavesdropping, they must and do adopt the
assumption that all errors in raw key are caused by Eve.

\bigskip\ 

As before, there are two stages to the protocol.

\subsection{Stage 1. Communication over a quantum channel}

\bigskip\ 

This stage is exactly the same as before, except that errors are now also
induced by noise.

\subsection{Stage 2. Communication in four phases over a public channel}

\bigskip\ 

In stage 2, Alice and Bob communicate over a public channel in four phases.
Phase 1 is dedicated to \textbf{raw key} extraction, phase 2 to \textbf{error
estimation}, phase 3 to \textbf{reconciliation}, i.e., to \textbf{reconciled
key} extraction, and phase 4 to \textbf{privacy amplification}, i.e.,
extraction of \textbf{final secret key}.

\subsubsection{Phase 1 of Stage 2. Extraction of raw key}

\bigskip\ 

This stage is the same as before, except Alice and Bob also delete those bit
locations at which Bob should have received but did not receive a bit. Such
``non-receptions'' could be caused by Eve's intrusion or by dark counts in
Bob's detecting device. The location of the dark counts are, of course,
communicated by Bob to Alice over the public channel.

\subsubsection{Phase 2 of Stage 2. Estimation of error in raw key}

\bigskip\ 

Alice and Bob now use the public channel to estimate the error rate in raw
key. They publicly select and agree upon a random sample of raw key, publicly
compare these bits to obtain an estimate $R$ of the error-rate. These revealed
bits are discarded from raw key. If $R$ exceeds a certain threshold $R_{Max}$,
then it will be impossible for Alice and Bob to arrive at a common secret key.
If so, Alice and Bob return to stage 1 to start over. On the other hand, If
the error estimate $R$ does not exceed $R_{Max}$, then Alice and Bob move onto
phase 3.

\subsubsection{Phase 3 of Stage 2. Extraction of reconciled key}

\bigskip\ 

In phase 3\footnote{The procedure given in Phase 3 Stage 2 is only one of many
possible procedures. \ In fact, there are now much more efficient procedures
than the procedure described below.}, Alice and Bob's objective is to remove
all errors from what remains of raw key to produce an error free common key,
called \textbf{reconciled key}. This phase is of course called
\textbf{reconciliation}, and takes place in two steps \cite{Bennett1} .

In step 1, Alice and Bob publicly agree upon a random permutation, and apply
it to what remains of their respective raw keys. Next Alice and Bob partition
the remnant raw key into blocks of length $\ell$, where the length $\ell$ is
chosen so that blocks of this length are unlikely to contain more than one
error. For each of these blocks, Alice and Bob publicly compare overall parity
checks, making sure each time to discard the last bit of the compared block.
Each time a overall parity check does not agree, Alice and Bob initiate a
binary search for the error, i.e., bisecting the block into two subblocks,
publicly comparing the parities for each of these subblocks, discarding the
right most bit of each subblock. They continue their bisective search on the
subblock for which their parities are not in agreement. This bisective search
continues until the erroneous bit is located and deleted. They then continue
to the next $\ell$-block.

Step 1 is repeated, i.e., a random permutation is chosen, remnant raw key is
partitioned into blocks of length $\ell$, parities are compared, etc. This is
done until it becomes inefficient to continue in this fashion.

Alice and Bob then move to step 2 by using a more refined reconciliation
procedure. They publicly select randomly chosen subsets of remnant raw key,
publicly compare parities, each time discarding an agreed upon bit from their
chosen key sample. If a parity should not agree, they employ the binary search
strategy of step 1 to locate and delete the error.

Finally, when, for some fixed number $N$ of consecutive repetitions of step 2,
no error is found, Alice and Bob assume that to a very high probability, the
remnant raw key is without error. Alice and Bob now rename the remnant raw
key\textbf{\ reconciled key}, and move on to the final and last phase of their communication.

\subsubsection{Phase 4 of Stage 2. Privacy amplification, i.e., extraction of
final secret key}

\bigskip\ 

Alice and Bob now have a common reconciled key which they know is only
partially secret from Eve. They now begin the process of \textbf{privacy
amplification}, which is the extraction of a secret key from a partially
secret one \cite{Bennett1} \cite{Bennett8}.

Based on their error estimate $R$, Alice and Bob obtain an upper bound $k$ of
the number of bits known by Eve of their $n$ bits of reconciled key. Let $s$
be a security parameter that Alice and Bob adjust as desired. They then
publicly select $n-k-s$ random subsets of reconciled key, without revealing
their contents, and without revealing their parities. The undisclosed parities
become the common \textbf{final secret key}. It can be shown that Eve's
average information about the final secret key is less than $2^{-s}/\ln2$ bits.

\bigskip\ 

\subsection{``Priming the pump'' to start authentication}

Unfortunately, there is no known way to initiate authentication without
initially exchanging secret key over a secure communication channel. So,
quantum protocols have not entirely overcome the ``catch 22'' of classical
cryptography. However, this secret key exchange for authentication need only
be done once. Thereafter, a portion of the secure key communicated via a
quantum protocol can be used for authentication.

\bigskip\ 

\section{The B92 quantum cryptographic protocol}

As with the BB84 quantum protocol, the B92 protocol \cite{Bennett2} can be
described in terms of any quantum system represented by a two dimensional
Hilbert space. For our description, we choose the two dimensional Hilbert
space $\mathcal{H}$ representing the polarization states of a single photon.

B92 can be implemented in terms of any non-orthogonal basis. We choose as our
non-orthogonal basis the kets
\[
\left|  \theta\right\rangle \text{ \quad and\quad}\left|  \overline{\theta
}\right\rangle ,
\]
where $\left|  \theta\right\rangle $ and $\left|  \overline{\theta
}\right\rangle $ denote respectively the kets representing the polarization
state of a photon linearly polarized at an angle $\theta$ and an angle
$-\theta$ with respect to the vertical, where $0<\theta<\pi/4$.

Unlike BB84 which requires two orthogonal quantum alphabets, B92 requires only
a single non-orthogonal quantum alphabet. We choose the non-orthogonal quantum
alphabet $\mathcal{A}_{\theta}$:
\[%
\begin{tabular}
[c]{c||c}%
Symbol & Bit\\\hline\hline
$\left|  \theta\right\rangle $ & $1$\\\hline
$\left|  \overline{\theta}\right\rangle $ & $0$\\\hline
& \\
\multicolumn{2}{c}{Linear Polarization}\\
\multicolumn{2}{c}{Quantum Alphabet $\mathcal{A}_{\theta}$}%
\end{tabular}
\]

\bigskip\ 

As in BB84, Alice and Bob communicate in two stages, the first over a one-way
quantum channel, the second over a two-way public channel.

\bigskip\ 

\subsection{Stage 1. Communication over a quantum channel}

Alice uses the quantum alphabet $\mathcal{A}_{\theta}$ to send her random
binary sequence to Bob. Since $\left|  \theta\right\rangle $ and $\left|
\overline{\theta}\right\rangle $ are not orthogonal, there is no one
experiment that will unambiguously distinguish between these two polarization states.

\bigskip\ 

Bob can use one of many possible measurement strategies. Bennett
\cite{Bennett2} suggests the measurements be based on the two incompatible
experiments corresponding to the projection operators
\[
P_{\lnot\theta}=1-\left|  \theta\right\rangle \left\langle \theta\right|
\text{\quad and\quad}P_{\lnot\overline{\theta}}=1-\left|  \overline{\theta
}\right\rangle \left\langle \overline{\theta}\right|
\]
In this case, Bob either correctly detects Alice's transmitted bit, or an
ambiguous result, i.e., an \textbf{erasure}, denoted by ``$?$''. Assuming that
Alice transmits 0's and 1's at random with equal probability and that, for
each incoming bit, Bob at random with equal probability chooses to base his
experiment on either of the incompatible operators $P_{\lnot\theta}$ or
$P_{\lnot\overline{\theta}}$, then the probability of Bob's correctly
receiving Alice's transmission is
\[
\frac{1-\left\|  \left\langle \theta\mid\overline{\theta}\right\rangle
\right\|  ^{2}}{2}%
\]
and the probability of receiving an erasure is
\[
\frac{1+\left\|  \left\langle \theta\mid\overline{\theta}\right\rangle
\right\|  ^{2}}{2}%
\]
where
\[
\left\|  \left\langle \theta\mid\overline{\theta}\right\rangle \right\|
=\cos\left(  2\theta\right)
\]
and where $0<\theta<\pi/4$. Thus, Bob receives more than 50\% erasures.

\bigskip\ 

On the other hand, Ekert et al \cite{Ekert1} suggest a more efficient
measurement process for Bob. They suggest that Bob base his experiments on the
\textbf{positive operator valued measure} (\textbf{POVM}) \cite{Busch1}
\cite{Peres2} consisting of the operators
\[
A_{\theta}=\frac{P_{\lnot\theta}}{1+\left\|  \left\langle \theta\mid
\overline{\theta}\right\rangle \right\|  }\text{,\quad}A_{\overline{\theta}%
}=\frac{P_{\lnot\overline{\theta}}}{1+\left\|  \left\langle \theta
\mid\overline{\theta}\right\rangle \right\|  }\text{, and\quad}A_{?}%
=1-A_{\theta}-A_{\overline{\theta}}%
\]
With this more efficient detection method, the probability of an inconclusive
result is now
\[
\left\|  \left\langle \theta\mid\overline{\theta}\right\rangle \right\|
=\cos\left(  2\theta\right)
\]
where again $0<\theta<\pi/4$.

\bigskip\ 

\subsubsection{Stage 2. Communication in four phases over a public channel}

Stage2 for the B92 protocol is the same as that for the BB84 protocol except
for phase 1.

\bigskip\ 

In phase 1 of stage 2, Bob publicly informs Alice as to which time slots he
received non-erasures. The bits in these time slots become Alice's and Bob's
raw keys.

Eve's presence is detected by an unusual error rate in Bob's raw key. It is
also possible to detect Eve's presence by an unusual erasure rate for Bob.
However, Ekert et al \cite{Ekert1} do point out that Eve can choose
eavesdropping strategies which have no effect on the erasure rate, and hence,
can only be detected by unusual error rates in Bob's raw key\footnote{This is
true for all 2-state protocols. On the other hand, for $n$-state protocols
with $n>2$, Eve's presence is always detectable from rejected key. See section
7 of this paper.}.

\bigskip\ 

\section{EPR quantum cryptographic protocols}

Ekert in \cite{Ekert6} has devised a quantum protocol based on the properties
of quantum-correlated particles.

\bigskip\ 

Einstein, Podolsky, and Rosen (EPR) in the their famous 1935 paper
\cite{Einstein1} challenged the foundations of quantum mechanics by pointing
out a ``paradox.'' There exist spatially separated pairs of particles,
henceforth called \textbf{EPR pairs}, whose states are correlated in such a
way that the measurement of a chosen observable $A$ of one automatically
determines the result of the measurement of $A$ of the other. Since EPR pairs
can be pairs of particles separated at great distances, this leads to what
appears to be a paradoxical ``action at a distance.''

For example, it is possible to create a pair of photons (each of which we
label below with the subscripts 1 and 2, respectively) with correlated linear
polarizations. An example of such an entangled state is given by
\[
\left|  \Omega_{0}\right\rangle =\frac{1}{\sqrt{2}}\left(  \left|
0\right\rangle _{1}\left|  \frac{\pi}{2}\right\rangle _{2}-\left|  \frac{\pi
}{2}\right\rangle _{1}\left|  0\right\rangle _{2}\right)
\]
where the notation $\left|  \theta\right\rangle $ has been defined in the
previous section. Thus, if one photon is measured to be in the vertical linear
polarization state $\left|  0\right\rangle $, the other, when measured, will
be found to be in the horizontal linear polarization state $\left|
\pi/2\right\rangle $, and vice versa.

Einstein et al \cite{Einstein1} then state that such quantum correlation
phenomena could be a strong indication that quantum mechanics is incomplete,
and that there exist ``hidden variables,'' inaccessible to experiments, which
explain such ``action at a distance.''

In 1964, Bell \cite{Bell1} gave a means for actually testing for
\textbf{locally hidden variable} (\textbf{LHV}) theories. He proved that all
such LHV theories must satisfy the \textbf{Bell inequality}. Quantum mechanics
has been shown to violate the inequality.

\bigskip\ 

The \textbf{EPR quantum protocol} is a 3-state protocol that uses Bell's
inequality to detect the presence or absence of Eve as a hidden variable.
Following the theme of this paper, we now describe this protocol in terms of
the polarization states of an EPR photon pair. As the three possible
polarization states of our EPR pair, we choose
\[%
\begin{tabular}
[c]{l}%
$\left|  \Omega_{0}\right\rangle =\frac{1}{\sqrt{2}}\left(  \left|
0\right\rangle _{1}\left|  \frac{3\pi}{6}\right\rangle _{2}-\left|  \frac
{3\pi}{6}\right\rangle _{1}\left|  0\right\rangle _{2}\right)  $.\\
\\
$\left|  \Omega_{1}\right\rangle =\frac{1}{\sqrt{2}}\left(  \left|  \frac{\pi
}{6}\right\rangle _{1}\left|  \frac{4\pi}{6}\right\rangle _{2}-\left|
\frac{4\pi}{6}\right\rangle _{1}\left|  \frac{\pi}{6}\right\rangle
_{2}\right)  $, and\\
\\
$\left|  \Omega_{2}\right\rangle =\frac{1}{\sqrt{2}}\left(  \left|  \frac
{2\pi}{6}\right\rangle _{1}\left|  \frac{5\pi}{6}\right\rangle _{2}-\left|
\frac{5\pi}{6}\right\rangle _{1}\left|  \frac{2\pi}{6}\right\rangle
_{2}\right)  $%
\end{tabular}
\]

\bigskip\ 

For each of these states, we choose the following corresponding mutually
non-orthogonal alphabets $\mathcal{A}_{0}$, $\mathcal{A}_{1}$,and
$\mathcal{A}_{2}$, given by the following tables:
\[%
\begin{tabular}
[c]{c||c}%
Symbol & Bit\\\hline\hline
$\left|  0\right\rangle $ & $0$\\\hline
$\left|  \frac{3\pi}6\right\rangle $ & $1$\\\hline
& \\
\multicolumn{2}{c}{Linear Polarization}\\
\multicolumn{2}{c}{Quantum Alphabet $\mathcal{A}_{0}$}%
\end{tabular}
\text{\quad}
\begin{tabular}
[c]{c||c}%
Symbol & Bit\\\hline\hline
$\left|  \frac\pi6\right\rangle $ & $0$\\\hline
$\left|  \frac{4\pi}6\right\rangle $ & $1$\\\hline
& \\
\multicolumn{2}{c}{Linear Polarization}\\
\multicolumn{2}{c}{Quantum Alphabet $\mathcal{A}_{1}$}%
\end{tabular}
\text{\quad}
\begin{tabular}
[c]{c||c}%
Symbol & Bit\\\hline\hline
$\left|  \frac{2\pi}6\right\rangle $ & $0$\\\hline
$\left|  \frac{5\pi}6\right\rangle $ & $1$\\\hline
& \\
\multicolumn{2}{c}{Linear Polarization}\\
\multicolumn{2}{c}{Quantum Alphabet $\mathcal{A}_{2}$}%
\end{tabular}
\]
The corresponding measurement operators chosen for these alphabets are
respectively
\[
\mathcal{M}_{0}=\left|  0\right\rangle \left\langle 0\right|  \text{,\quad
}\mathcal{M}_{1}=\left|  \frac\pi6\right\rangle \left\langle \frac\pi6\right|
\text{, and }\mathcal{M}_{2}=\left|  \frac{2\pi}6\right\rangle \left\langle
\frac{2\pi}6\right|
\]

\bigskip\ 

As with the BB84 and B92 , there are two stages to the EPR protocol, the first
stage over a quantum channel, the second over a public channel.

\bigskip\ 

\subsection{Stage 1. Communication over a quantum channel}

For each time slot, a state $\left|  \Omega_{j}\right\rangle $ is randomly
selected with equal probability from the set of states $\left\{  \left|
\Omega_{0}\right\rangle ,\left|  \Omega_{1}\right\rangle ,\left|  \Omega
_{2}\right\rangle \right\}  $. Than an EPR pair is created in the selected
state $\left|  \Omega_{j}\right\rangle $. One photon of the constructed EPR
pair is sent to Alice, the other to Bob. Alice and Bob at random with equal
probability separately and independently select one of the three measurement
operators $\mathcal{M}_{0}$, $\mathcal{M}_{1}$, and $\mathcal{M}_{2}$, and
accordingly measure their respective photons. Alice records her measured bit.
On the other hand, Bob records the complement of his measured bit. This
procedure is repeated for as many time slots as needed.

\bigskip\ 

\subsection{Stage 2. Communication over a public channel}

In stage 2, Alice and Bob communicate over a public channel.

\subsubsection{Phase 1 of Stage2. Separation of key into raw and rejected keys}

In phase 1 of stage 2, Alice and Bob carry on a discussion over a public
channel to determine those bit slots at which they used the same measurement
operators. They each then separate their respective bit sequences into two
subsequences. One subsequence, called \textbf{raw key}, consists of those bit
slots at which they used the same measurement operators. The other
subsequence, called \textbf{rejected key}, consists of all the remaining bit slots.

\bigskip\ 

\subsubsection{Phase 2 of Stage 2. Detection of Eve's presence with Bell's
inequality applied to rejected key}

Unlike the BB84 and B92 protocols, the EPR protocol, instead of discarding
rejected key, actually uses it to detect Eve's presence. Alice and Bob now
carry on a discussion over a public channel comparing their respective
rejected keys to determine whether or not Bell's inequality is satisfied. If
it is, Eve's presence is detected. If not, then Eve is absent.

\bigskip\ 

For the EPR protocol, Bell's inequality can be written as follows. Let
$P\left(  \neq\mid i,j\right)  $ denote the probability that two corresponding
bits of Alice's and Bob's rejected keys do not match given that the
measurement operators chosen by Alice and Bob are respectively either
$\mathcal{M}_{i}$ and $\mathcal{M}_{j}$ or $\mathcal{M}_{j}$ and
$\mathcal{M}_{i}$. Let $P\left(  =\mid i,j\right)  =1-P\left(  \neq\mid
i,j\right)  $. Let
\[
\Delta\left(  i,j\right)  =P\left(  \neq\mid i,j\right)  -P\left(  =\mid
i,j\right)
\]
Finally, let
\[
\beta=1+\Delta\left(  1,2\right)  -\left|  \Delta\left(  0,1\right)
-\Delta\left(  0,2\right)  \right|
\]

\bigskip\ 

Then Bell's inequality in this case reduces to
\[
\beta\geq0
\]
Moreover, for quantum mechanics (i.e., no hidden variables)
\[
\beta=-\frac12
\]
which is a clear violation of Bell's inequality.

\bigskip\ 

\subsubsection{Phase 3 of Stage 2. Reconciliation}

In the presence of noise, the remaining phase of the EPR protocol is
reconciliation, as described in the BB84 and B92 protocols.

\bigskip\ 

\section{Other protocols}

It is not possible to cover all possible quantum protocols in this paper.
There is the EPR protocol with a single particle. There is also a 2-state EPR
implementation of the BB84 protocol. For details, see \cite{Bennett7}
\cite{D'Espagnat1}. For various \textbf{multiple state} and \textbf{rejected
data protocols}, see \cite{Blow1}.

\bigskip\ 

\section{Eavesdropping strategies and counter measures}

There are many eavesdropping strategies available to Eve. (See for example
\cite{Ekert1},\cite{Brandt2}.) We list only a few.

\bigskip\ 

\subsection{Opaque eavesdropping}

For this strategy, Eve intercepts Alice's message, and then masquerades as
Alice by sending her received message on to Bob. Opaque eavesdropping has
already been discussed in sections 4 and 5 of this paper. For more
information, the reader is referred to \cite{Ekert1}.

\bigskip\ 

\subsection{Translucent eavesdropping without entanglement}

For this strategy, Eve makes the information carrier interact unitarily with
her probe, and then lets it proceed on to Bob in a slightly modified state. In
the case of the B92 protocol, Eve's detection probe with initial state
$\left|  \Psi\right\rangle $ would perform a unitary transformation $U$ of the
form
\[
\left\{
\begin{array}
[c]{c}%
\left|  \theta\right\rangle \left|  \Psi\right\rangle \mapsto U\left|
\theta\right\rangle \left|  \Psi\right\rangle =\left|  \theta^{\prime
}\right\rangle \left|  \Psi_{\theta}\right\rangle \\
\\
\left|  \overline{\theta}\right\rangle \left|  \Psi\right\rangle \mapsto
U\left|  \overline{\theta}\right\rangle \left|  \Psi\right\rangle =\left|
\overline{\theta}^{\prime}\right\rangle \left|  \Psi_{\overline{\theta}%
}\right\rangle
\end{array}
\right.
\]
where $\left|  \theta^{\prime}\right\rangle $ and $\left|  \overline{\theta
}^{\prime}\right\rangle $ denote the slightly changed states received by Bob
after the action of the probe, and where $\left|  \Psi_{\theta}\right\rangle $
and $\left|  \Psi_{\overline{\theta}}\right\rangle $ denote the states of the
probe after the transformation.. We refer the reader to \cite{Ekert1} for an
in depth analysis of this eavesdropping strategy.

\bigskip\ 

\subsection{Translucent eavesdropping with entanglement}

For this strategy, Eve entangles the state of her probe and the carrier, and
then she sends the carrier on to Bob. In the case of the B92 protocol, Eve's
detection probe with initial state $\left|  \Psi\right\rangle $ would perform
a unitary transformation $U$ of the form
\[
\left\{
\begin{array}
[c]{c}%
\left|  \theta\right\rangle \left|  \Psi\right\rangle \mapsto U\left|
\theta\right\rangle \left|  \Psi\right\rangle =a\left|  \theta\right\rangle
\left|  \Psi_{\theta}\right\rangle +b\left|  \overline{\theta}\right\rangle
\left|  \Psi_{\overline{\theta}}\right\rangle \\
\\
\left|  \overline{\theta}\right\rangle \left|  \Psi\right\rangle \mapsto
U\left|  \overline{\theta}\right\rangle \left|  \Psi\right\rangle =b\left|
\theta\right\rangle \left|  \Psi_{\theta}\right\rangle +a\left|
\overline{\theta}\right\rangle \left|  \Psi_{\overline{\theta}}\right\rangle
\end{array}
\right.
\]
We refer the reader to \cite{Ekert1}, \cite{Brandt2} for an in depth analysis
of this eavesdropping strategy.

\bigskip\ 

\subsection{Countermeasures to Eve's eavesdropping strategies}

As far as the author has been able to determine, all quantum intrusion
detection algorithms in the open literature depend on some assumption as to
which eavesdropping strategy is chosen by Eve. It is important that
eavesdropping algorithms be developed that detect Eve's intrusion no matter
which eavesdropping strategy she chooses to use. (For some insight in
intrusion detection algorithms, the reader is referred to \cite{Ekert1}%
,\cite{Brandt2}.)

\bigskip\ 

\section{Conclusion}

It is not easy to emphasize how dramatic an impact the application of quantum
mechanics has had and will have on cryptographic communication systems. From
the perspective of defensive cryptography, it is now within the realm of
possibility to build practical cryptographic systems which check for, detect,
and prevent unauthorized intrusion. Quantum mechanics provides an intrusion
detection mechanism never thought possible within the world of classical
cryptography. Most importantly, the feasibility of these methods has been
experimentally verified in a laboratory setting.

Moreover, from the perspective of offensive cryptography, the application of
quantum mechanics to computation also holds forth the promise of a dramatic
increase of computational parallelism for cryptanalytic attacks. Shor's
quantum factoring algorithm \cite{Shor1} \cite{Ekert3} is just one example of
such potential. However, unlike quantum protocols, quantum computational
parallelism has yet to be fully verified in a laboratory setting.

\bigskip\ 

Much remains to be done before quantum cryptography is a truly practical and
useful tool for cryptographic communication. We list below some of the areas
in need of development:

\begin{itemize}
\item  Quantum protocols need to be extended to a computer network setting.
(See \cite{Phoenix3} and \cite{Townsend5}.)

\item  More sophisticated error correction and detection techniques need to be
implemented in quantum protocols. (See \cite{Bennett1}, \cite{Bennett8}, and
\cite{Bennett13}.)

\item  There is a need for greater understanding of intrusion detection in the
presence of noise. The no cloning theorem of Appendix A of this paper and the
``no detection implies no information'' theorem of Appendix B of this paper
simply do not provide a complete picture. (See \cite{Ekert1}.)

\item  There is a need for better intrusion detection algorithms. As far as
the author has been able to determine, all quantum intrusion detection
algorithms in the open literature depend on some assumption as to which
eavesdropping strategy is chosen by Eve. It is important that eavesdropping
algorithms be developed that detect Eve's intrusion no matter which
eavesdropping strategy she uses. (See \cite{Ekert1}.)
\end{itemize}

\bigskip

\section{Acknowledgment}

I would like to thank Howard Brandt for his helpful discussions, and the
referees for their helpful suggestions. Finally I would like to thank Alan
Sherman for his encouragement to publish this paper.\medskip

\section{Addendum}

Quantum cryptography has continued its rapid pace of development since this
paper was written. There is the recent experimental work found in
\cite{Muller1}, \cite{Muller2}. Progress has been made in correcting errors
received from noisy channels \cite{Briegel1}, \cite{Briegel2}, \cite{Enk1},
\cite{Enk2}. A number of protocols, in particular, the quantum bit commitment
protocol, have been shown to be insecure \cite{Lo1}, \cite{Lo2},
\cite{Mayers1}. There has been progress in the development of multi-user
quantum cryptography \cite{Townsend6}. The security of quantum cryptography
against collective key attacks has been studied \cite{Biham1}. There have been
at least two independent claims of the proof of ultimate security of quantum
cryptography, i.e., security against all possible attacks \cite{Lo3},
\cite{Mayers2}, \cite{Mayers3}, \cite{Mayers4}. Finally, although tangentially
related to this paper, it should be mentioned that a new quantum algorithm for
searching databases has been developed \cite{Grover1}, \cite{Grover2},
\cite{Grover3}. \pagebreak 

\section{Appendix A. The no cloning theorem}

In this appendix, we prove that there can be no device that produces exact
replicas or copies of a quantum system. If such a ``quantum copier'' existed,
then Eve could eavesdrop without detection. This proof is taken from
\cite{Peres2}. It is an amazingly simple application of the linearity of
quantum mechanics. (See also \cite{Wootters1} for a proof using the creation
operators of quantum electrodynamics.) \medskip\ 

Let us assume that there exists a \textbf{quantum replicator} initially in
state $\left|  \Psi\right\rangle $ which duplicates quantum systems via a
unitary transformation $U$.\medskip\ 

Let $\left|  u\right\rangle $ and $\left|  v\right\rangle $ be two arbitrary
states such that
\[
0<\left\|  \left\langle u\mid v\right\rangle \right\|  <1.
\]
Then the application of the quantum replicator to $\left|  u\right\rangle $
and $\left|  v\right\rangle $ yields
\[%
\begin{array}
[c]{c}%
\left|  \Psi\right\rangle \left|  u\right\rangle \mapsto U\left|
\Psi\right\rangle \left|  u\right\rangle =\left|  \Psi^{\prime}\right\rangle
\left|  u\right\rangle \left|  u\right\rangle \\
\\
\left|  \Psi\right\rangle \left|  v\right\rangle \mapsto U\left|
\Psi\right\rangle \left|  u\right\rangle =\left|  \Psi^{\prime\prime
}\right\rangle \left|  v\right\rangle \left|  v\right\rangle
\end{array}
\]
where $\left|  \Psi^{\prime}\right\rangle $ and $\left|  \Psi^{\prime\prime
}\right\rangle $ denote the states of the quantum replicator after the two
respective duplications.\medskip\ 

Thus,
\[
\left\langle u\right|  \left\langle \Psi\right|  U^{\dagger}U\left|
\Psi\right\rangle \left|  v\right\rangle =\left\langle u\right|  \left\langle
\Psi\mid\Psi\right\rangle \left|  v\right\rangle =\left\langle u\mid
v\right\rangle ,
\]
because of the unitarity of $U$ and because $\left\langle \Psi\mid
\Psi\right\rangle =1$. On the other hand,
\[
\left\langle u\right|  \left\langle u\right|  \left\langle \Psi^{\prime}%
\mid\Psi^{\prime\prime}\right\rangle \left|  v\right\rangle \left|
v\right\rangle =\left\langle \Psi^{\prime}\mid\Psi^{\prime\prime}\right\rangle
\left\langle u\mid v\right\rangle ^{2}\text{.}%
\]
As a result, we have the equation
\[
\left\langle u\mid v\right\rangle =\left\langle \Psi^{\prime}\mid\Psi
^{\prime\prime}\right\rangle \left\langle u\mid v\right\rangle ^{2}%
\]
But this equation cannot be satisfied since $\left\|  \left\langle
\Psi^{\prime}\mid\Psi^{\prime\prime}\right\rangle \right\|  \leq1$ and
$\left|  u\right\rangle $ and $\left|  v\right\rangle $ were chosen so that
$0<\left\|  \left\langle u\mid v\right\rangle \right\|  <1$.\medskip\ 

Hence, a quantum replicator cannot exist. \pagebreak 

\section{Appendix B. Proof that an undetectable eavesdropper can obtain no
information from the B92 protocol}

In this appendix we prove that an undetectable eavesdropper for the B92
protocol obtains no information whatsoever. The proof is taken from
\cite{Bennett7}.\medskip\ 

Let $\left|  a\right\rangle $ and $\left|  b\right\rangle $ denote the two
non-orthogonal states used in the B92 protocol\footnote{In section 6 of this
paper we denoted these states by $\left|  \theta\right\rangle $ and $\left|
\overline{\theta}\right\rangle $.}. Thus,
\[
\left\langle a\mid b\right\rangle \neq0
\]
Let $U$ be the unitary transformation performed by Eve's detection probe,
which we assume is initially in state $\left|  \Psi\right\rangle $.\medskip\ 

Since Eve's probe is undetectable, we have
\[%
\begin{array}
[c]{c}%
\left|  \Psi\right\rangle \left|  a\right\rangle \mapsto U\left|
\Psi\right\rangle \left|  a\right\rangle =\left|  \Psi^{\prime}\right\rangle
\left|  a\right\rangle \\
\\
\left|  \Psi\right\rangle \left|  b\right\rangle \mapsto U\left|
\Psi\right\rangle \left|  b\right\rangle =\left|  \Psi^{\prime\prime
}\right\rangle \left|  b\right\rangle
\end{array}
\]
where $\left|  \Psi^{\prime}\right\rangle $ and $\left|  \Psi^{\prime\prime
}\right\rangle $ denote the states of Eve's prober after the detection of
$\left|  a\right\rangle $ and $\left|  b\right\rangle $ respectively. Please
note that, since Eve is undetectable, her probe has no effect on the states
$\left|  a\right\rangle $ and $\left|  b\right\rangle $. So $\left|
a\right\rangle $ appears on both sides of the first equation, and $\left|
b\right\rangle $ appears on both sides of the second equation.

Thus,
\[
\left\langle a\right|  \left\langle \Psi\right|  U^{\dagger}U\left|
\Psi\right\rangle \left|  b\right\rangle =\left\langle a\right|  \left\langle
\Psi\mid\Psi\right\rangle \left|  b\right\rangle =\left\langle a\mid
b\right\rangle ,
\]
because of the unitarity of $U$ and because $\left\langle \Psi\mid
\Psi\right\rangle =1$. On the other hand,
\[
\left\langle a\right|  \left\langle \Psi^{\prime}\mid\Psi^{\prime\prime
}\right\rangle \left|  b\right\rangle =\left\langle \Psi^{\prime}\mid
\Psi^{\prime\prime}\right\rangle \left\langle a\mid b\right\rangle \text{.}%
\]
As a result, we have the equation
\[
\left\langle a\mid b\right\rangle =\left\langle \Psi^{\prime}\mid\Psi
^{\prime\prime}\right\rangle \left\langle a\mid b\right\rangle
\]
\medskip\ 

But $\left\langle a\mid b\right\rangle \neq0$ implies that $\left\langle
\Psi^{\prime}\mid\Psi^{\prime\prime}\right\rangle =1$. Since $\left|
\Psi^{\prime}\right\rangle $ and $\left|  \Psi^{\prime\prime}\right\rangle $
are normalized, this implies that $\left|  \Psi^{\prime}\right\rangle =\left|
\Psi^{\prime\prime}\right\rangle $. It follows that Eve's probe is in the same
state no matter which of the states $\left|  a\right\rangle $ and $\left|
b\right\rangle $ is received. Thus, Eve obtains no information whatsoever.
\pagebreak \ 

\section{Appendix C. Part of a Rosetta stone for quantum mechanics.}

\medskip

This appendix is intended for readers unfamiliar with quantum mechanics. It's
purpose is to provide those readers with enough background in quantum
mechanics to understand a substantial portion of this paper. Because of space
limitations, this appendix is of necessity far from a complete overview of the
subject. \medskip\ 

\subsection{Polarized light: Part I. The classical perspective}

Light waves in the vacuum are transverse electromagnetic (EM) waves with both
electric and magnetic field vectors perpendicular to the direction of
propagation and also to each other. (See figure 6.) \medskip

\qquad\qquad%
{\parbox[b]{4.0914in}{\begin{center}
\includegraphics[
height=1.9536in,
width=4.0914in
]%
{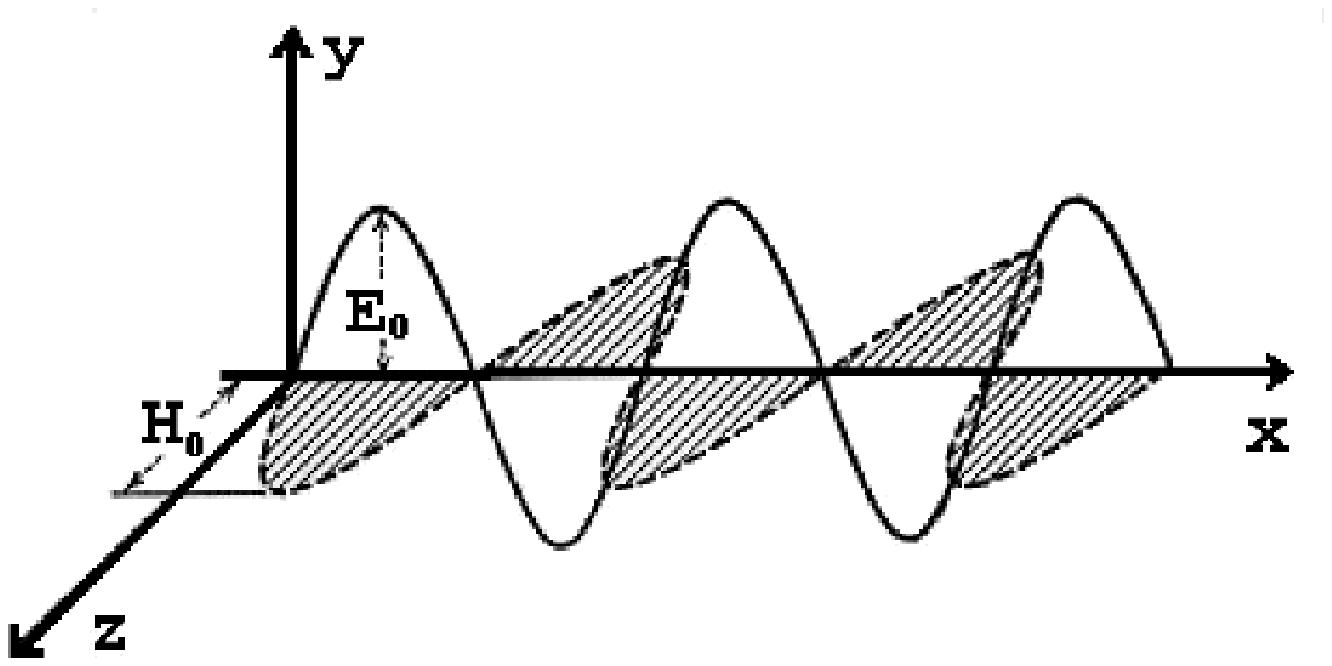}%
\\
Figure 6. A linearly polarized electromagnetic wave.
\end{center}}}%
\medskip

\noindent If the electric field vector is always parallel to a fixed line,
then the EM wave is said to be \textbf{linearly polarized}. If the electric
field vector rotates about the direction of propagation forming a
right-(left-)handed screw, it is said to be \textbf{right} (\textbf{left})
\textbf{elliptically} \textbf{polarized}. If the rotating electric field
vector inscribes a circle, the EM wave is said to be right-or
left-\textbf{circularly polarized}.\medskip\ 

\subsection{A Rosetta stone for Dirac notation:\ Part I. Bras, kets, and
bra-(c)-kets\medskip\ }

A \textbf{Hilbert space} $\mathcal{H}$ is a vector space over the complex
numbers $\mathbb{C}$ with a complex valued inner product
\[
\left(  -,-\right)  :\mathcal{H}\times\mathcal{H\rightarrow}\mathbb{C}%
\]
which is complete with respect to the norm
\[
\left\|  u\right\|  =\sqrt{\left(  u,u\right)  }%
\]
induced by the inner product. \medskip\ 

\begin{remark}
By a complex valued inner product, we mean a map
\[
\left(  -,-\right)  :\mathcal{H}\times\mathcal{H\rightarrow}\mathbb{C}%
\]
from $\mathcal{H\times H}$ into the complex numbers $\mathbb{C}$ such that:

\begin{enumerate}
\item [\qquad1)]$\left(  u,u\right)  =0$ if and only if $u=0$

\item[\qquad2)] $(u,v)=(v,u)^{*}$

\item[\qquad3)] $(u,v+w)=(u,v)+(u,w)$

\item[\qquad4)] $(u,\lambda v)=\lambda(u,v)$
\end{enumerate}

where `$^{*}$' denotes the complex conjugate.
\end{remark}

\begin{remark}
(Please note that $(\lambda u,v)=\lambda^{*}(u,v)$. )
\end{remark}

\medskip\ The elements of $\mathcal{H}$ will be called \textbf{ket vectors},
\textbf{state kets}, or simply \textbf{kets}. They will be denoted as:
\[
\left|  \,label\,\right\rangle
\]
where `$label$' denotes some label.\medskip\ 

Let $\mathcal{H}^{\#}$ denote the Hilbert space of all Hilbert space morphisms
of $\mathcal{H}$ into the Hilbert space of all complex numbers $\mathbb{C}$,
i.e.,
\[
\mathcal{H}^{\#}=Hom_{\mathbb{C}}\left(  \mathcal{H},\mathbb{C}\right)
\text{.}%
\]
The elements of $\mathcal{H}^{\#}$ will be called \textbf{bra vectors},
\textbf{state bras}, or simply \textbf{bras}. They will be denoted as:
\[
\left\langle \,label\,\right|
\]
where once again `$label$' denotes some label.\medskip\ 

Also please note that the complex number
\[
\left\langle \,label_{1}\,\right|  \left(  \left|  \,label_{2}\,\right\rangle
\right)
\]
will simply be denoted by
\[
\left\langle \,label_{1}\,\mid\,label_{2}\,\right\rangle
\]
and will be called the \textbf{bra-(c)-ket} product of the bra $\left\langle
\,label_{1}\,\right|  $ and the ket $\left|  \,label_{2}\,\right\rangle
$.\medskip\ 

There is a monomorphism (which is an isomorphism if the underlying Hilbert
space is finite dimensional)
\[
\mathcal{H}\overset{\#}{\rightarrow}\mathcal{H}^{\#}%
\]
defined by
\[
\left|  \,label\,\right\rangle \longmapsto\left(  \;\left|
\,label\,\right\rangle ,-\right)
\]
The bra $\left(  \;\left|  \,label\,\right\rangle ,-\right)  $ is denoted by
$\left\langle \,label\,\right|  $.\medskip\ 

Hence,
\[
\left\langle \,label_{1}\,\mid\,label_{2}\,\right\rangle =\left(  \left|
\,label_{1}\,\right\rangle ,\left|  \,label_{2}\,\right\rangle \right)
\]
\medskip\ 

\begin{remark}
Please note that $\left(  \lambda\left|  \,label\,\right\rangle \right)
^{\#}=\lambda^{*}\left\langle label\right|  $.\medskip\ 
\end{remark}

The \textbf{tensor product}\footnote{Readers well versed in homological
algebra will recognize this informal definition as a slightly disguised
version of the more rigorous universal definition of the tensor product. For
more details, please refer to \cite{Cartan1}, or any other standard reference
on homological algebra.} $\mathcal{H}\otimes\mathcal{K}$ of two Hilbert spaces
$\mathcal{H}$ and $\mathcal{K}$ is simply the ``simplest'' Hilbert space such that

\begin{description}
\item [\qquad1)]$\left(  h_{1}+h_{2}\right)  \otimes k=h_{1}\otimes
k+h_{2}\otimes k$, for all $h$, $h_{1}$, $h_{2}\in\mathcal{H}$ and for all $k
$, $k_{1}$, $k_{2}\in\mathcal{K}$, and

\item[\qquad2)] $h\otimes\left(  k_{1}+k_{2}\right)  =h\otimes k_{1}+h\otimes
k_{2}$ for all $h$, $h_{1}$, $h_{2}\in\mathcal{H}$ and for all $k$, $k_{1}$,
$k_{2}\in\mathcal{K}$.
\end{description}

\noindent It immediately follows that

\begin{description}
\item [\qquad3)]$\lambda\left(  h\otimes k\right)  \equiv\left(  \lambda
h\right)  \otimes k=h\otimes\left(  \lambda k\right)  $ for all $\lambda
\in\mathbb{C}$, $h\in\mathcal{H}$, $k\in\mathcal{K}$.
\end{description}

Finally, if $\left|  \,label_{1}\,\right\rangle $ and $\left|  \,label_{2}%
\,\right\rangle $ are kets respectively in Hilbert spaces $\mathcal{H}_{1}$
and $\mathcal{H}_{2}$, then their tensor product will be written in any one of
the following three ways:
\[%
\begin{array}
[c]{c}%
\left|  \,label_{1}\,\right\rangle \otimes\left|  \,label_{2}\,\right\rangle
\\
\\
\left|  \,label_{1}\,\right\rangle \left|  \,label_{2}\,\right\rangle \\
\\
\left|  \,label_{1}\,,\,label_{2}\,\right\rangle
\end{array}
\]
$\,$\medskip

\subsection{Polarized light: Part II. The quantum mechanical perspective}

The states of a quantum mechanical system are represented by state kets in a
Hilbert space $\mathcal{H}$. Two kets $\left|  \alpha\right\rangle $ and
$\left|  \beta\right\rangle $ represent the same quantum mechanical state if
they differ by a non-zero multiplicative constant. I.e., $\left|
\alpha\right\rangle $ and $\left|  \beta\right\rangle $ represent the same
quantum mechanical state if there exists a non-zero $\lambda\in\mathbb{C}$
such that
\[
\left|  \alpha\right\rangle =\lambda\left|  \beta\right\rangle
\]
Hence, the quantum mechanical states are the elements of the manifold
\[
\mathcal{H}/\symbol{126}=\mathbb{C}P^{n}%
\]
where $n$ denotes the dimension of $\mathcal{H}$, and $\mathbb{C}P^{n}$
denotes complex projective space.

\begin{description}
\item [Convention:]Since a quantum mechanical state is represented by a state
ket up to a multiplicative constant, we will unless stated otherwise, choose
those kets $\left|  \alpha\right\rangle $ which are unit normal, i.e., such
that
\[
\left\langle \alpha\mid\alpha\right\rangle =1\Longleftrightarrow\left\|
\,\left|  \alpha\right\rangle \right\|  =1
\]
\medskip\ 
\end{description}

The polarization states of a photon are represented as state kets in a two
dimensional Hilbert space $\mathcal{H}$. One orthonormal basis of
$\mathcal{H}$ consists of the kets
\[
\left|  \curvearrowleft\right\rangle \text{ and }\left|  \curvearrowright
\right\rangle
\]
which represent respectively the quantum mechanical states of left- and
right-circularly polarized photons. Another orthonormal basis consists of the
kets
\[
\left|  \updownarrow\right\rangle \text{ and }\left|  \leftrightarrow
\right\rangle
\]
representing respectively vertically and horizontally linearly polarized
photons. And yet another orthonormal basis consists of the kets
\[
\left|  \nearrow\right\rangle \text{ and }\left|  \searrow\right\rangle
\]
for linearly polarized photons at the angles $\theta=\pi/4$ and $\theta
=-\pi/4$ off the vertical, respectively.\medskip\ 

These orthonormal bases are related as follows:
\[
\left\{
\begin{array}
[c]{ccc}%
\left|  \nearrow\right\rangle  & = & \frac{1}{\sqrt{2}}\left(  \left|
\updownarrow\right\rangle +\left|  \leftrightarrow\right\rangle \right) \\
&  & \\
\left|  \searrow\right\rangle  & = & \frac{1}{\sqrt{2}}\left(  \left|
\updownarrow\right\rangle -\left|  \leftrightarrow\right\rangle \right)
\end{array}
\right.  \qquad\qquad\qquad\left\{
\begin{array}
[c]{ccc}%
\left|  \nearrow\right\rangle  & = & \frac{1+i}{2}\left|  \curvearrowright
\right\rangle +\frac{1-i}{2}\left|  \curvearrowleft\right\rangle \\
&  & \\
\left|  \searrow\right\rangle  & = & \frac{1-i}{2}\left|  \curvearrowright
\right\rangle +\frac{1+i}{2}\left|  \curvearrowleft\right\rangle
\end{array}
\right.
\]
\medskip\ %

\[
\left\{
\begin{array}
[c]{ccc}%
\left|  \updownarrow\right\rangle  & = & \frac{1}{\sqrt{2}}\left(  \left|
\nearrow\right\rangle +\left|  \nwarrow\right\rangle \right) \\
&  & \\
\left|  \leftrightarrow\right\rangle  & = & \frac{1}{\sqrt{2}}\left(  \left|
\nearrow\right\rangle -\left|  \nwarrow\right\rangle \right)
\end{array}
\right.  \qquad\qquad\qquad\left\{
\begin{array}
[c]{ccc}%
\left|  \updownarrow\right\rangle  & = & \frac{1}{\sqrt{2}}\left(  \left|
\curvearrowright\right\rangle +\left|  \curvearrowleft\right\rangle \right) \\
&  & \\
\left|  \leftrightarrow\right\rangle  & = & \frac{i}{\sqrt{2}}\left(  \left|
\curvearrowright\right\rangle -\left|  \curvearrowleft\right\rangle \right)
\end{array}
\right.
\]
\medskip\ %

\[
\left\{
\begin{array}
[c]{ccc}%
\left|  \curvearrowright\right\rangle  & = & \frac{1}{\sqrt{2}}\left(  \left|
\updownarrow\right\rangle -i\left|  \leftrightarrow\right\rangle \right) \\
&  & \\
\left|  \curvearrowleft\right\rangle  & = & \frac{1}{\sqrt{2}}\left(  \left|
\updownarrow\right\rangle +i\left|  \leftrightarrow\right\rangle \right)
\end{array}
\right.  \qquad\qquad\qquad\left\{
\begin{array}
[c]{ccc}%
\left|  \curvearrowright\right\rangle  & = & \frac{1-i}{2}\left|
\nearrow\right\rangle +\frac{1+i}{2}\left|  \nwarrow\right\rangle \\
&  & \\
\left|  \curvearrowleft\right\rangle  & = & \frac{1+i}{2}\left|
\nearrow\right\rangle +\frac{1-i}{2}\left|  \nwarrow\right\rangle
\end{array}
\right.
\]
\medskip

The bracket products of the various polarization kets are given in the table
below:\
\[%
\begin{tabular}
[c]{|c||c|c||c|c||c|c|}\hline
& $\left|  \updownarrow\right\rangle $ & $\left|  \leftrightarrow\right\rangle
$ & $\left|  \nearrow\right\rangle $ & $\left|  \nwarrow\right\rangle $ &
$\left|  \curvearrowright\right\rangle $ & $\left|  \curvearrowleft
\right\rangle $\\\hline\hline
$\left\langle \updownarrow\right|  $ & $1$ & $0$ & $\frac{1}{\sqrt{2}}$ &
$\frac{1}{\sqrt{2}}$ & $\frac{1}{\sqrt{2}}$ & $\frac{1}{\sqrt{2}}$\\\hline
$\left\langle \leftrightarrow\right|  $ & $0$ & $1$ & $\frac{1}{\sqrt{2}}$ &
$-\frac{1}{\sqrt{2}}$ & $-\frac{i}{\sqrt{2}}$ & $\frac{i}{\sqrt{2}}%
$\\\hline\hline
$\left\langle \nearrow\right|  $ & $\frac{1}{\sqrt{2}}$ & $\frac{1}{\sqrt{2}}
$ & $1$ & $0$ & $\frac{1+i}{2}$ & $\frac{1-i}{2}$\\\hline
$\left\langle \nwarrow\right|  $ & $\frac{1}{\sqrt{2}}$ & $-\frac{1}{\sqrt{2}}
$ & $0$ & $1$ & $\frac{1-i}{2}$ & $\frac{1+i}{2}$\\\hline\hline
$\left\langle \curvearrowright\right|  $ & $\frac{1}{\sqrt{2}}$ & $\frac
{i}{\sqrt{2}}$ & $\frac{1-i}{2}$ & $\frac{1+i}{2}$ & $1$ & $0$\\\hline
$\left\langle \curvearrowleft\right|  $ & $\frac{1}{\sqrt{2}}$ & $-\frac
{i}{\sqrt{2}}$ & $\frac{1+i}{2}$ & $\frac{1-i}{2}$ & $0$ & $1$\\\hline
\end{tabular}
\]
\medskip\ 

\subsection{A Rosetta stone for Dirac notation:\ Part II. Operators\medskip\ }

An \textbf{(linear) operator} or \textbf{transformation} $\mathcal{O}$ on a
ket space $\mathcal{H}$ is a Hilbert space morphism of $\mathcal{H}$ into
$\mathcal{H}$, i.e., is an element of
\[
Hom_{\mathbb{C}}\left(  \mathcal{H},\mathcal{H}\right)
\]
\medskip\ 

The \textbf{adjoint} $\mathcal{O}^{\dagger}$ of an operator $\mathcal{O}$ is
that operator such that
\[
\left(  \mathcal{O}^{\dagger}\left|  \,label_{1}\,\right\rangle ,\left|
\,label_{2}\,\right\rangle \right)  =\left(  \left|  \,label_{1}%
\,\right\rangle ,\mathcal{O}\left|  \,label_{2}\,\right\rangle \right)
\]
for all kets $\left|  \,label_{1}\,\right\rangle $ and $\left|  \,label_{2}%
\,\right\rangle $.\medskip\ 

In like manner, an (linear) operator or transformation on a bra space
$\mathcal{H}^{\#}$ is an element of
\[
Hom_{\mathbb{C}}\left(  \mathcal{H}^{\#},\mathcal{H}^{\#}\right)
\]
Moreover, each operator $\mathcal{O}$ on $\mathcal{H}$ can be identified with
an operator, also denoted by $\mathcal{O}$, on $\mathcal{H}^{\#}$ defined by
\[
\left\langle \,label_{1}\,\right|  \longmapsto\left\langle \,label_{1}%
\,\right|  \mathcal{O}%
\]
where $\left\langle \,label_{1}\,\right|  \mathcal{O}$ is the bra defined by
\[
\left(  \left\langle \,label_{1}\,\right|  \mathcal{O}\right)  \left(  \left|
\,label_{2}\right\rangle \right)  =\left\langle \,label_{1}\,\right|  \left(
\mathcal{O}\left|  \,label_{2}\right\rangle \right)
\]
(This is sometimes called Dirac's associativity law.) Hence, the expression
\[
\left\langle \,label_{1}\,\right|  \mathcal{O}\left|  \,label_{2}%
\right\rangle
\]
is unambiguous.\medskip\ 

\begin{remark}
Please note that
\[
\left(  \mathcal{O}\left|  \,label\right\rangle \right)  ^{\#}=\left\langle
label\right|  \mathcal{O}^{\dagger}%
\]
\medskip\ 
\end{remark}

In quantum mechanics, an \textbf{observable} is simply a \textbf{Hermitian}
(also called \textbf{self-adjoint}) operator on a Hilbert space $\mathcal{H}$,
i.e., an operator $\mathcal{O}$ such that
\[
\mathcal{O}^{\dagger}=\mathcal{O}\text{ .}%
\]
An \textbf{eigenvalue} $a$ of an operator $A$ is a complex number for which
there is a ket $\left|  label\right\rangle $ such that
\[
A\left|  label\right\rangle =a\left|  label\right\rangle \text{ .}%
\]
The ket $\left|  label\right\rangle $ is called an \textbf{eigenket} of $A$
corresponding to the eigenvalue $a$. \medskip\ 

An important theorem about observables is given below:

\begin{theorem}
The eigenvalues $a_{i}$ of an observable $A$ are all real numbers. Moreover,
the eigenkets for distinct eigenvalues of an observable are
orthogonal.\bigskip\ 
\end{theorem}

\begin{definition}
An eigenvalue is \textbf{degenerate} if there are at least two linearly
independent eigenkets for that eigenvalue. Otherwise, it is
\textbf{nondegenerate}.\medskip\ 
\end{definition}

\begin{description}
\item [Notational Convention:]If all the eigenvalues $a_{i}$ of an observable
$A$ are nondegenerate, then we can and do label the eigenkets of $A$ with the
eigenvalues $a_{i}$. Thus, we can write:
\[
A\left|  a_{i}\right\rangle =a_{i}\left|  a_{i}\right\rangle
\]
for each eigenvalue $a_{i}$. In this paper, unless stated otherwise, we assume
that the eigenvalues of observables are non-degenerate.\medskip\ 
\end{description}

One exception to the above notational convention is the \textbf{measurement
operator}
\[
\left|  a_{i}\right\rangle \left\langle a_{i}\right|
\]
for the eigenvalue $a_{i}$, which is the outer product of ket $\left|
a_{i}\right\rangle $ with its adjoint $\left\langle a_{i}\right|  $. It has
two eigenvalues $0$ and $1$. $1$ is a nondegenerate eigenvalue with eigenket
$\left|  a_{i}\right\rangle $. $0$ is a degenerate eigenvalue with
corresponding eigenkets $\left\{  \,\left|  a_{j}\right\rangle \,\right\}
_{j\neq i}$ .\medskip\ 

An observable $A$ is said to be \textbf{complete} if its eigenkets $\left|
a_{i}\right\rangle $ form a basis (hence, an orthonormal basis) of the Hilbert
space $\mathcal{H}$. Given a complete nondegenerate observable $A$, then any
ket $\left|  \psi\right\rangle $ in $\mathcal{H}$ can be written as:
\[
\left|  \psi\right\rangle =\sum_{i}\left|  a_{i}\right\rangle \left\langle
a_{i}\mid\psi\right\rangle
\]
Thus, for a complete nondegenerate observable $A$, we have the following
operator equation which expresses the completeness of $A$,
\[
\sum_{i}\left|  a_{i}\right\rangle \left\langle a_{i}\right|  =1
\]
Thus, in this notation, we have
\[
A=\sum_{i}a_{i}\left|  a_{i}\right\rangle \left\langle a_{i}\right|
\]

\subsection{Quantum measurement: General principles}

In this section, $A$ will denote a complete nondegenerate observable with
eigenvalues $a_{i}$ and eigenkets $\left|  a_{i}\right\rangle $ . \medskip

According to quantum measurement theory, the measurement of an observable $A$
of a ket $\left|  \psi\right\rangle $ with respect to the basis $\left\{
\left|  a_{i}\right\rangle \right\}  $ produces the eigenvalue $a_{i}$ with
probability
\[
Prob\left(  \text{Value\quad}a_{i}\text{\quad is\quad observed}\right)
=\left\|  \left\langle a_{i}\mid\psi\right\rangle \right\|  ^{2}%
\]
and forces the state of the quantum system to become the corresponding
eigenket $\left|  a_{i}\right\rangle $.\medskip\ 

Since quantum measurement is such a hotly debated topic among physicists, we
(in self-defense) quote P.A.M. Dirac\cite{Dirac1}:

\begin{quote}
``A measurement always causes the (quantum mechanical) system to jump into an
eigenstate of the dynamical variable that is being measured.''\medskip\ 
\end{quote}

Thus, the above mentioned measurement of observable $A$ of ket $\left|
\psi\right\rangle $ can be diagrammatically represented as follows:
\[
\left|  \psi\right\rangle =\sum_{i}\left|  a_{i}\right\rangle \left\langle
a_{i}\mid\psi\right\rangle \quad%
\begin{array}
[c]{c}%
\text{Meas. of }A\\
\Longrightarrow\\
Prob=\left\|  \left\langle a_{j}\mid\psi\right\rangle \right\|  ^{2}%
\end{array}
\quad a_{j}\left|  a_{j}\right\rangle \quad\symbol{126}\quad\left|
a_{j}\right\rangle \quad%
\begin{array}
[c]{c}%
\text{Meas. of }A\\
\Longrightarrow\\
Prob=1
\end{array}
\quad\left|  a_{j}\right\rangle
\]
\ \medskip\ 

The observable
\[
\left|  a_{i}\right\rangle \left\langle a_{i}\right|
\]
is frequently called a \textbf{selective measurement operator} (or a
\textbf{filtration}) for $a_{i}$. As mentioned earlier, it has two eigenvalues
$0$ and $1$. $1$ is a nondegenerate eigenvalue with eigenket $\left|
a_{i}\right\rangle $, and $0$ is a degenerate eigenvalue with eigenkets
$\left\{  \left|  a_{j}\right\rangle \right\}  _{j\neq i}$.\medskip\ 

Thus,
\[
\left|  \psi\right\rangle \quad%
\begin{array}
[c]{c}%
\left|  a_{i}\right\rangle \left\langle a_{i}\right| \\
\Longrightarrow\\
Prob=\left\|  \left\langle a_{i}\mid\psi\right\rangle \right\|  ^{2}%
\end{array}
\quad1\cdot\left|  a_{i}\right\rangle =\left|  a_{i}\right\rangle \text{ ,}%
\]
\medskip\ but for $j\neq i$,
\[
\left|  \psi\right\rangle \quad%
\begin{array}
[c]{c}%
\left|  a_{i}\right\rangle \left\langle a_{i}\right| \\
\Longrightarrow\\
Prob=\left\|  \left\langle a_{j}\mid\psi\right\rangle \right\|  ^{2}%
\end{array}
\quad0\cdot\left|  a_{j}\right\rangle =0
\]
\medskip

\subsection{Polarized light: Part III. Three examples of quantum measurement\ }

We can now apply the above general principles of quantum measurement to
polarized light. Three examples are given below:\footnote{The last two
examples can easily be verified experimentally with at most three pair of
polarized sunglasses.}

\begin{example}%
\[%
\begin{array}
[c]{c}%
\text{Rt. Circularly}\\
\text{polarized photon}\\
\qquad\\
\left|  \curvearrowleft\right\rangle =\frac{1}{\sqrt{2}}\left(  \left|
\updownarrow\right\rangle +i\left|  \leftrightarrow\right\rangle \right) \\
\qquad
\end{array}
\qquad\Longrightarrow\qquad%
\begin{array}
[c]{c}%
\text{Vertical}\\
\text{Polaroid}\\
\text{filter}\\%
{\includegraphics[
trim=0.000000in 0.000000in 0.012478in 0.010417in,
height=0.6624in,
width=0.3918in
]%
{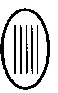}%
}%
\\
\text{Measurement op.}\\
\left|  \updownarrow\right\rangle \left\langle \updownarrow\right|
\end{array}
\qquad%
\begin{array}
[c]{ccc}%
&  & \text{Vertically}\\
&  & \text{polarized}\\
Prob=\frac{1}{2} &  & \text{photon}\\
\Longrightarrow &  & \left|  \updownarrow\right\rangle \\
&  & \\
&  & \\
\Longrightarrow &  & 0\\
Prob=\frac{1}{2} &  & \text{No photon}\\
&  & \\
&  &
\end{array}
\]
\medskip\ 
\end{example}

\begin{example}
A vertically polarized filter followed by a horizontally polarized filter.
\end{example}%

\[%
\begin{array}
[c]{c}%
\text{Entangled}\\
\text{photon}\\
\qquad\\
\alpha\left|  \updownarrow\right\rangle +\beta\left|  \leftrightarrow
\right\rangle \\
\qquad\\
\text{Normalized so that}\\
\left\|  \alpha\right\|  ^{2}+\left\|  \beta\right\|  ^{2}=1
\end{array}
\Longrightarrow%
\begin{array}
[c]{c}%
\text{Vert.}\\
\text{polar.}\\
\text{filter}\\%
{\includegraphics[
trim=0.000000in 0.000000in 0.012478in 0.010417in,
height=0.6624in,
width=0.3918in
]%
{vertical.ps}%
}%
\\
\qquad\\
\left|  \updownarrow\right\rangle \left\langle \updownarrow\right|
\end{array}
.
\begin{array}
[c]{c}%
Prob=\left\|  \alpha\right\|  ^{2}\\
\qquad\\
\Longrightarrow
\end{array}%
\begin{array}
[c]{c}%
\text{Vert.}\\
\text{polar.}\\
\text{photon}\\
\qquad\\
\left|  \updownarrow\right\rangle
\end{array}
\Longrightarrow%
\begin{array}
[c]{c}%
\text{Horiz.}\\
\text{polar.}\\
\text{filter}\\%
{\includegraphics[
trim=0.000000in 0.000000in 0.025593in 0.010417in,
height=0.6624in,
width=0.3918in
]%
{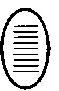}%
}%
\\
\qquad\\
\left|  \leftrightarrow\right\rangle \left\langle \leftrightarrow\right|
\end{array}
\quad%
\begin{array}
[c]{cc}%
& \text{No}\\
& \text{photon}\\
Prob=1 & \\
\Longrightarrow & 0\\
& \\
& \\
. &
\end{array}
\]
\medskip\ 

\begin{example}
But if we insert a diagonally polarized filter (by $45^{o}$ off the vertical)
between the two polarized filters in the above example, we have:
\[%
\begin{array}
[c]{c}%
\\
\\%
{\includegraphics[
trim=0.000000in 0.000000in 0.012478in 0.010417in,
height=0.6624in,
width=0.3918in
]%
{vertical.ps}%
}%
\\
\\
\left|  \updownarrow\right\rangle \left\langle \updownarrow\right|
\end{array}%
\begin{array}
[c]{c}%
\left\|  \alpha\right\|  ^{2}\\
\\
\Rightarrow
\end{array}
\left|  \updownarrow\right\rangle =\frac{1}{\sqrt{2}}\left(  \left|
\nearrow\right\rangle +\left|  \nwarrow\right\rangle \right)
\begin{array}
[c]{c}%
\\
\\%
{\includegraphics[
trim=0.000000in 0.000000in 0.025593in 0.000000in,
height=0.6607in,
width=0.3918in
]%
{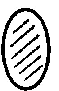}%
}%
\\
\\
\left|  \nearrow\right\rangle \left\langle \nearrow\right|
\end{array}%
\begin{array}
[c]{c}%
\frac{1}{2}\\
\\
\Rightarrow
\end{array}
\left|  \nearrow\right\rangle =\frac{1}{\sqrt{2}}\left(  \left|
\updownarrow\right\rangle +\left|  \leftrightarrow\right\rangle \right)
\begin{array}
[c]{c}%
\\
\\%
{\includegraphics[
trim=0.000000in 0.000000in 0.025593in 0.010417in,
height=0.6624in,
width=0.3918in
]%
{horizon.ps}%
}%
\\
\\
\left|  \leftrightarrow\right\rangle \left\langle \leftrightarrow\right|
\end{array}%
\begin{array}
[c]{c}%
\frac{1}{2}\\
\\
\Rightarrow
\end{array}
\left|  \leftrightarrow\right\rangle
\]
\medskip\ 
\end{example}

where the input to the first filter is $\alpha\left|  \updownarrow
\right\rangle +\beta\left|  \leftrightarrow\right\rangle $.\medskip\ 

\subsection{A Rosetta stone for Dirac notation: Part III. Expected values}

The \textbf{average value} (\textbf{expected value}) of a measurement of an
observable $A$ on a state $\left|  \alpha\right\rangle $ is:
\[
\left\langle A\right\rangle =\left\langle \alpha\right|  A\left|
\alpha\right\rangle
\]
For, since
\[
\sum_{i}\left|  a_{i}\right\rangle \left\langle a_{i}\right|  =1\text{ ,}%
\]
we have%

\[
\left\langle A\right\rangle =\left\langle \alpha\right|  A\left|
\alpha\right\rangle =\left\langle \alpha\right|  \left(  \sum_{i}\left|
a_{i}\right\rangle \left\langle a_{i}\right|  \right)  A\left(  \sum
_{j}\left|  a_{j}\right\rangle \left\langle a_{j}\right|  \right)  \left|
\alpha\right\rangle =\sum_{i,j}\left\langle \alpha\mid a_{i}\right\rangle
\left\langle a_{i}\right|  A\left|  a_{j}\right\rangle \left\langle a_{j}%
\mid\alpha\right\rangle
\]
But on the other hand,
\[
\left\langle a_{i}\right|  A\left|  a_{j}\right\rangle =a_{j}\left\langle
a_{i}\mid a_{j}\right\rangle =a_{i}\delta_{ij}%
\]
Thus,
\[
\left\langle A\right\rangle =\sum_{i}\left\langle \alpha\mid a_{i}%
\right\rangle a_{i}\left\langle a_{i}\mid\alpha\right\rangle =\sum_{i}%
a_{i}\left\|  \left\langle a_{i}\mid\alpha\right\rangle \right\|  ^{2}%
\]
Hence, we have the standard expected value formula,
\[
\left\langle A\right\rangle =\sum_{i}a_{i}Prob\left(  \text{Observing }%
a_{j}\text{ on input }\left|  \alpha\right\rangle \right)
\]
\medskip

\subsection{Dynamics of closed quantum systems: Unitary transformations, the
Hamiltonian, and Schr\"{o}dinger's equation}

\bigskip

An operator $U$ on a Hilbert space $\mathcal{H}$ is \textbf{unitary} if
\[
U^{\dagger}=U^{-1}\text{ .}%
\]
Unitary operators are of central importance in quantum mechanics for many
reason. We list below only two:

\begin{itemize}
\item  Closed quantum mechanical systems transform only via unitary transformations

\item  Unitary transformations preserve quantum probabilities
\end{itemize}

Let $\left|  \psi(t)\right\rangle $ denote the state of a closed quantum
mechanical system $\mathcal{S}$ as a function of time $t$. Then the dynamical
behavior of $\mathcal{S}$ is determined by the \textbf{Schr\"{o}dinger
equation}
\[
\frac{\partial}{\partial t}\left|  \psi(t)\right\rangle =-\frac{i}{\hslash
}H\left|  \psi(t)\right\rangle \text{ ,}%
\]
and boundary conditions, where $\hslash$ denotes Planck's constant and $H$
denotes an observable of $\mathcal{S}$ called the \textbf{Hamiltonian}. The
Hamiltonian is the quantum mechanical analog of the Hamiltonian classical
mechanics. In classical physics, it is the total energy of the system.\medskip

\subsection{There is much more to quantum mechanics}

There is much more to quantum mechanics. For more in-depth overviews, there
are many outstanding books. Among such books are \cite{Feynman1},
\cite{Sakurai1}, \cite{Dirac1}, \cite{Omnes1}, \cite{Peres2}, and many more.
Some excellent insights into this subject are also given in chapter 2 of
\cite{Penrose1}.

\pagebreak

\end{document}